\documentclass[aps,prb,twocolumn,groupedaddress]{revtex4}

\usepackage{graphicx}
\usepackage{amsmath}

\begin{document}

\title{Tunneling spectroscopy and Majorana modes emergent from topological gapless phases in high-$T_c$ cuprate superconductors}
\author{Jun-Ting Kao$^{1}$, Shin-Ming Huang$^{2}$, Chung-Yu Mou$^{1,3,4}$, and C. C. Tsuei$^{5}$}
\affiliation{$^{1}$Department of Physics, National Tsing Hua University, Hsinchu 30043,
Taiwan}
\affiliation{$^{2}$Graphene Research Centre and Department of Physics, National University of Singapore, Singapore 117542}
\affiliation{$^{3}$Institute of Physics, Academia Sinica, Nankang, Taiwan}
\affiliation{$^{4}$Physics Division, National Center for Theoretical Sciences, P.O. Box
2-131, Hsinchu, Taiwan}
\affiliation{$^{5}$IBM Thomas J. Watson Research Center, Yorktown Heights, NY 10598, U.S.A.}

\begin{abstract}
We explore possible signatures for observing Majorana fermions in the tunneling spectroscopy of high-$T_c$ superconductors. It is shown that due to the Rashba spin-orbit interaction ($\alpha_R$) generally 
introduced by contact electrode,  in addition to the Heisenberg spin exchange interaction, the Dzyaloshinskii-Moriya and spin dipole-dipole interactions are induced. As a result, $p$-wave superconductivity is induced with the gap function $d$-vector being not aligned with the internal magnetic field of the spin-orbit interaction. For typical strength of the Rashba interaction, the induced $p$-wave is weak. Hence the resulting superconductor is still gapless and is not a topological superconductor.  However, we find that the ground state undergoes a phase transition to a topological gapless phase with each nodal point originated from pure $d$-wave being split into two stable nodal points characterized by the symmetry class DIII. Due to the splitting nodal structure, zero-energy Majorana modes always exist for any interfaces that are not exactly in $(100)$ or $(010)$ directions. Hence for general interfaces, existence of Majorana modes is a robust feature. In addition, due to the non-aligned $d$-vector, for $(110)$ interfaces in which  $d$-wave is subdominant to $p$-wave,  there exist sizable dispersive Majorana edge states. Our results indicate that due to the presence of these Majorana modes, a small plateau in tunneling spectrum near zero bias peak would be induced. Furthermore, zero-energy Majorana modes result in 4$\pi$ periodicity in typical SIS$'$  junctions with difference in orientations of S and S$'$ being within $21^\circ-39^\circ$ for $\alpha_R=0.05$ eV$-0.3$ eV. As a result, it is easy for a $\pi$-ring in tricrystal experiments to hold Majorana fermions and exhibit periods of two flux quanta in external magnetic field.  These phenomena may have been already observed in experiments and their connections to experimental results are discussed.
\end{abstract}

\date{\today}

\pacs{73.20.-r, 73.20.At,73.21.Hb}

\maketitle

\section{Introduction}
In the past decades, searching for the mysterious particle, Majorana fermion,  proposed by Ettore Majorana in 1937 \cite{Majorana} has continued to be an important issue in high-energy physics. Recently, the field of topological superconductors and superfluids\cite{Ludwig} in condensed matter physics has been drawing much attention since these topological matters are capable of hosting Majorana fermions.  Instead of seeking for it in high-energy experiments, topological matters in condensed matter systems may shed light on finding this elusive Majorana fermion. Majorana fermion, half state of a fermion, obeys non-Abelian statistics and possesses the non-local property. Both of the significant ingredients render such topological matters a great potential to realize quantum computers $-$ the dream which scientists have been pursuing for over 30 years. 

Topological matters are classified by distinct topological numbers based on 
dimension and symmetries of the system. 
In analogous to the framework of topological insulators characterized by a $Z_2$ invariance, non-centrosymmetric superconductors which break inversion symmetry but preserve time reversal symmetry are generally termed as time-reversal-invariant (TRI) superconductors. 
Inversion symmetry breaking leads to the mixing of spin-singlet and spin-triplet Cooper pairings\cite{Gorkov}. As a result, intensive studies have been dedicated to investigate
fully gapped TRI superconductors for their potential to be a topological nontrivial matter, termed as TRI topological superconductors\cite{KaneSC, Bernevig}. In order to have fully gapped superconductors, 
many proposals for creating Majorana fermions rely on the proximity effect of $s$-wave superconductors\cite{Beenakker}. The disadvantage of these approaches lies in the fact that most
$s$-wave superconductors have low critical temperatures and Majorana fermions can be
generated only in very low temperatures. It is therefore desirable to seek alternative approaches 
based on high-$T_c$ cuprate superconductors. 

In real materials, the heavy-fermion superconductor, $\rm{CePt_3Si}$ \cite{Sigrist1,Sigrist2}, discovered in the early days and the recent finding of superconductivity at the interface between $\rm{LaAlO_3}$ and $\rm{SrTiO_3}$\cite{Reyren} are typical examples of TRI superconductors in which spin-orbit interaction comes into play due to lack of inversion center. Recently, it is argued that
the contact of heavy metals with superconductors can induce the Rashba spin-orbit interaction in the interface\cite{PALee}. In particular, for Au/YBCO interface, it is estimated that the Rashba spin splitting energy at Fermi surface is 
of the order 200 meV\cite{Tsuei}. There is also experimental evidence showing that there may exist intrinsic spin-orbit interaction in cuprate superconductors\cite{Harrison}. From theoretical point of view, however, the strength of available spin orbit interaction is too small to turn a $d$-wave cuprate superconductor into a fully gapped  TRI superconductor. Therefore, even though the ideal scheme is to generate Majarona fermions via TRI topological superconductors, its feasibility is low in high $T_c$ cuprates and it is necessary to seek for alternative approaches. 

In this paper, we investigate effects of the Rashba spin-orbit interaction on high-$T_c$ cuprate superconductors.  Based on the mean-field theory of \it{t-J} \rm model\cite{Wen, Mou}, we show that the gap function $d$-vector for the induced spin-triplet pairing is generally not aligned with the internal magnetic field of the spin-orbit interaction. This fact respects the $C_{4v}$ point group symmetry and for typical strength of the Rashba interaction, the superconductor is still gapless with each nodal point of the original $d$-wave superconducting state being split.  The resulting nodal points are protected by symmetries in the symmetry class DIII\cite{Ryu}. Furthermore, due to the splitting nodal structures, zero-energy Majorana edge modes that are determined by projection of bulk nodes to edges\cite{DHLee} always exist for any orientations of interfaces that are not exactly in $[100]$ or $[010]$ directions. In particular, near $[110]$ direction in which  $d$-wave is subdominant to $p$-wave,  there exist sizable dispersive Majorana edge states. Our results indicate that the presence of these Majorana modes gives rises to a small plateau in tunneling spectrum near zero bias peak. Furthermore, these Majorana modes always result in 4$\pi$-Josephson effect in typical junctions and hence flux trapped in tricrystal experiments jump in unit of two flux quanta.  These phenomena may have been already observed in experiments and their connections to other experiments will be discussed.

This paper is organized as follows. In Sec.~\ref{model}, by including the Rashba spin-orbit interaction, we derive superconducting phase diagram based on the mean-field theory of the effective low-energy Hamiltonian for the strong coupling limit of the Hubbard model. In Sec.~\ref{gapless},  the gapless regime of $d+p$ wave resulted from typical strength of Rashba interaction is investigated. 
Edge states and their relations to nodal point structures are examined. In particular, how the edge states of $(110)$ edge change in the presence of the Rashba spin-orbit interaction changes is also examined.  In Sec.~\ref{tunnel}, we examine the tunneling spectrum for the NIS junction and periodicity of Josephson current in a SIS$'$ junction. The contribution due to the dispersive and flat-band Majorana modes are examined. Furthermore, the relation of 4$\pi$ periodicity and flat-band Majorana modes are derived for  Josephson junctions. The conditions for the $\pi$-ring in the tricrystal experiment to hold Majorana fermions are also examined in details. Finally, in Sec.~\ref{conclusion}, we conclude and discuss possible connections of our results to experimental observations. 

\begin{figure}[thp]
\includegraphics[width=4.5cm]{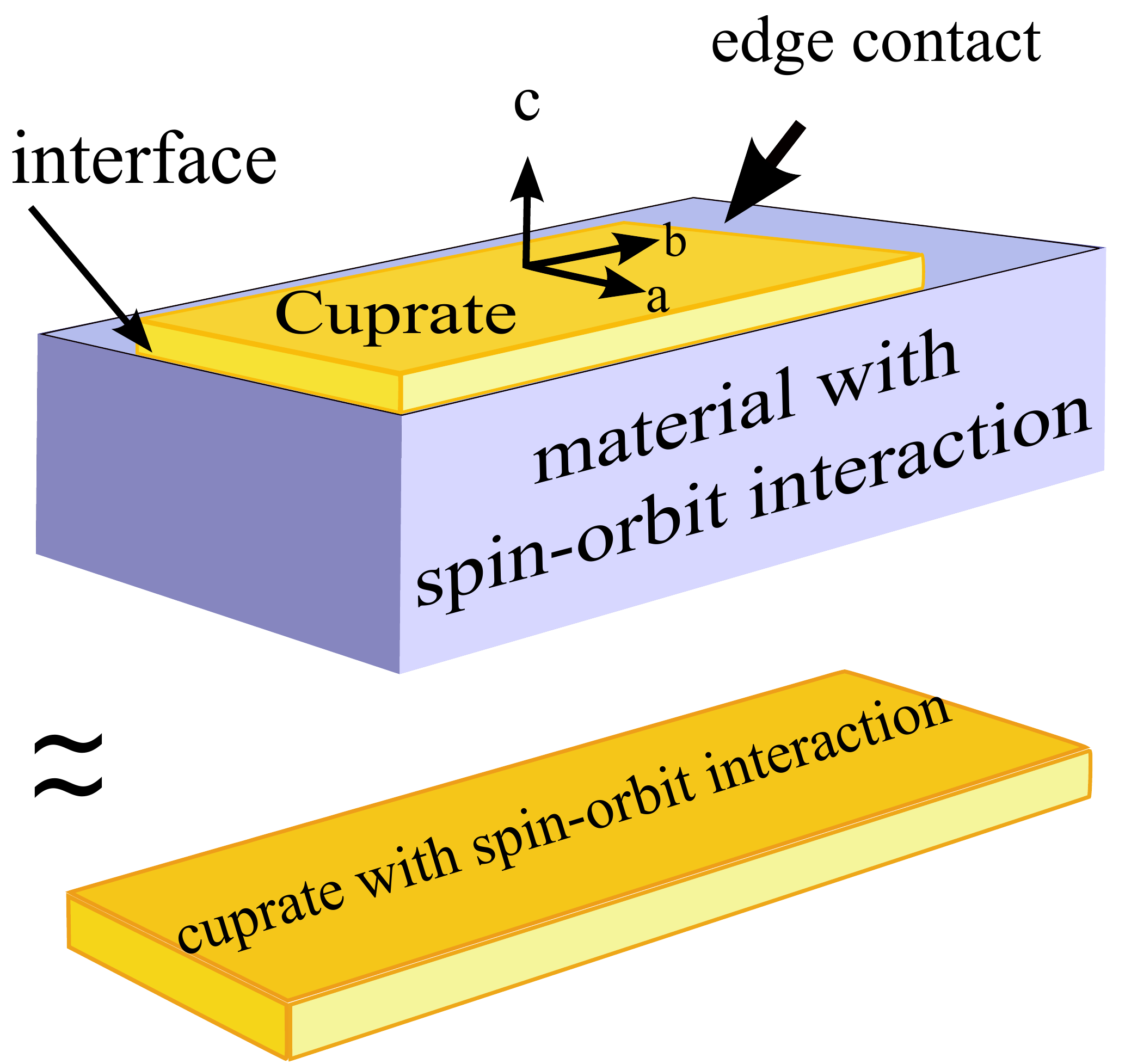}\\
\caption{Typical configuration of high-$T_c$ cuprate superconductors in contact with other materials (electrode or substrate). Through the interface between the cuprate and the underlying material, spin-orbit interactions that break inversion symmetry are induced in the cuprate superconductor either through proximity effect or in the layer near the interface.
Here the cuprate superconductor can make further tunnel junction with other superconductors via the edge contact.}
\label{Fig1}
\end{figure}

\section{Theoretical Model and Mean-Field Phase Diagram}
\label{model}

The parent compound of high-$T_c$ cuprate superconductor is well-known to be an antiferromagnetic Mott insulator. In spite of the fact that a complete theory for high-$T_c$ superconductors is under debate, a common consensus for grasping the essential physics behind the emergence of superconductivity lies in the 2D Hubbard model within large-\it{U} \rm limit, or the equivalent \it{t-J} \rm model\cite{Wen}. However, in typical experiments, high-$T_c$ cuprate superconductors are often in the form of thin films with contact to other materials along $c$-axis as illustrated in Fig.~\ref{Fig1}. Through the interface between the cuprate and the underlying material, spin-orbit interactions that break inversion symmetry may be induced either in the cuprate superconductor by proximity effect or induced in the layer near the interface.  In particular, for Au/YBCO interface, it is estimated that the energy splitting due to the spin-orbit interaction is of the order 0.2 eV\cite{Tsuei}. The largest splitting reported is also around 0.3eV\cite{BiTeI}. It is therefore necessary to include effects due to the spin-orbit interaction up to the spitting energy being around 0.3eV.

In the presence of the Rashba spin-orbit interaction, however, there is no full understanding of corresponding superconducting phases. From the symmetry point of view, the Rashba spin-orbit interaction breaks the inversion symmetry. As a result,
parity is not conserved and hence it is generally true that singlet pairing will be mixed with the triplet pairing in any superconducting phases. To be explicit, we shall start from the 2D Hubbard model with the Rashba spin-orbit interaction.
The effect of dispersion along $c$-axis will be examined at the end of Sec. IV.
In the strong coupling limit, exchange interactions give arise an extended $t-J$  model which can be generally expressed as $H_{eff}=H_t+H_J$ with
\begin{eqnarray}
\nonumber H_t &=& - \sum_{ij}t_{ij} \tilde{c}^\dagger_{i,\sigma}\tilde{c}_{j,\sigma} \\
&+& \sum_{<ij>} i\frac{\alpha_R}{2 a}\left[\mathbf{\sigma}\times (\mathbf{r}_j-\mathbf{r}_i) \right]_{\alpha \beta} \cdot\hat{z} (\tilde{c}^\dagger_{i,\alpha}\tilde{c}_{j,\beta}) +H.c.,  \nonumber \\
\\
\nonumber H_J&=&H_{J_1}+H_{J_2}+H_{J_3},\\
\nonumber&=& \sum_{<ij>}J_1\left(\mathbf{S}_i\cdot \mathbf{S}_j-\frac{1}{4}n_in_j\right)+J_2\left[\mathbf{D} \cdot\left(\mathbf{S}_i\times \mathbf{S}_j\right)\right],\\
&+&J_3\sum_{<ij>}\left[\mathbf{S}_{i}\cdot\mathbf{S}_{j}+\frac{1}{4}n_in_j-2(\mathbf{D} \cdot \mathbf{S}_i)(\mathbf{D} \cdot \mathbf{S}_j)\right]. \nonumber \\ 
\end{eqnarray}
Here $a$ is the lattice constant. $t_{ij}=t$, $t'$, and $t''$ for sites $i$ and $j$ are
nearest, next nearest, and the third nearest neighbors,
respectively, and $t_{ij}=0$ for longer distance. $\alpha_R$ characterizes the energy strength of 
the Rashba spin-orbit interaction.
$\tilde{c}_{i,\sigma} = (1-n_{i,-\sigma})c_{i,\sigma}$ satisfies
the no-double-occupancy constraint. For electrons
In terms of the on-site Coulomb interaction (Hubbard $U$), three spin interactions are given by  $J_1=\frac{4t^2}{U}$, $J_2=-\frac{4t\alpha_R}{U}$ and $J_3=-\frac{\alpha^2_R}{U}$, representing the Heisenberg interaction, the Dzyaloshinskii-Moriya (DM) interaction and the spin dipole-dipole interaction respectively. The orbital vector $\mathbf{D}$ that couples to spins is related to the lattice constant $a$ connecting sites in nearest neighbors along $x$($y$) direction by $\mathbf{D}=((\mathbf{r}_j-\mathbf{r}_i)_y,-(\mathbf{r}_j-\mathbf{r}_i)_x,0)\equiv(a,-a,0)$.

To satisfy the no-double-occupancy constraint, one can resort to  the slave-boson method\cite{Wen} or  adopt Gutzwiller approximations by using renormalized parameters\cite{renormalize}. In low doping regime, both approaches yield the same mean-field Hamiltonian.  We shall adopt the Gutzwiller approximation. In the Gutzwiller approximations, strength of spin interactions remain the same in
the low doping regime. Furthermore, $\tilde{c}_{i,\sigma}$ is replaced by $c_{i,\sigma}$.
The spin interactions can be generally factorized as summations of particle-particle channels and particle-hole channels\cite{Ubbens}. To investigate superconducting phases, one focuses on particle-particle channels. We find that  the $J_1$ term yields the spin-singlet pairing channel.  By contrast, the $J_3$ term contains the spin-triplet pairing channel, and the $J_2$ term mixes the spin-singlet and spin-triplet pairing channels. Specifically, the pairing channels can be written as 
\begin{eqnarray}
H_{J_1}&=&-\frac{J_1}{2}\sum_{\langle ij \rangle}\hat{\psi}^{\dagger}_{ij}\hat{\psi}_{ji},\\
H_{J_2}&=&J_2\sum_{<ij>}(\widetilde{\mathbf{D}}_{ij}^*\cdot\vec{\mathbf{\phi}}_{ij}^{\dagger})\hat{\psi}_{ji}+\hat{\psi}_{ij}^{\dagger}(\widetilde{\mathbf{D}}_{ij}\cdot\mathbf{\vec{\phi}}_{ji}),\\
H_{J_3}&=&J_3\sum_{\langle ij \rangle}\vec{\mathbf{\Phi}}_{ij}^{\dagger}\vec{\mathbf{\Phi}}_{ji},
\end{eqnarray}
where the singlet pairing field $\hat{\psi}_{ji}=c_{j\downarrow}c_{i\uparrow}-c_{j\uparrow}c_{i\downarrow}$, the spin-triplet vector field $\mathbf{\vec{\phi}^{\dagger}}_{ij}=(c^{\dagger}_{i\uparrow}c^{\dagger}_{j\uparrow},c^{\dagger}_{i\downarrow}c^{\dagger}_{j\downarrow},0)$ and $\vec{\mathbf{\Phi}}_{ij}^{\dagger}\equiv\frac{1}{\sqrt{2}}(c^{\dagger}_{i\uparrow}c^{\dagger}_{i+a \hat{x} \uparrow}+c^{\dagger}_{i\downarrow}c^{\dagger}_{i+a \hat{x} \downarrow},c^{\dagger}_{i\uparrow}c^{\dagger}_{i+a \hat{y}\uparrow}-c^{\dagger}_{i\downarrow}c^{\dagger}_{i+a \hat{y} \downarrow},0)$. The vector $\widetilde{\mathbf{D}}_{ij}$ are given by $\widetilde{\mathbf{D}}_{ii+x}=-\frac{a}{4}(1,1,0)$ and $\widetilde{\mathbf{D}}_{ii+y}=i\frac{a}{4}(-1,1,0)$.
\begin{table}
\label{table:C4}
\centering
\caption{ The representations of $C_{4v}$}
\begin{tabular}{|c| c c| c c|}
\hline
   & &   Even parity &   &Odd parity \\\hline
$A_1$& $A_{1g}$ & 1 & $A_{2u}$ & $k_x\hat{y}-k_y\hat{x};$\\
&   &   &   & $k_xk_y(k_x\hat{x}-k_y\hat{y});$\\
&   &   &   & $k_xk_yk_z(k^2_x-k^2_y)\hat{z}$ \\\hline 
$A_2$ & $A_{2g}$ & $k_xk_y(k^2_x-k^2_y)$ & $A_{1u}$ & $k_z\hat{z};k_x\hat{x}+k_y\hat{y}$  \\ \hline 
$B_1$ & $B_{1g}$ & $k^2_x-k^2_y$ & $B_{2u}$ &$k_x\hat{y}+k_y\hat{x}$  \\\hline 
$B_2$ & $B_{2g}$ & $k_xk_y$ & $B_{1u}$ & $k_x\hat{x}-k_y\hat{y}$ \\\hline 
$E$ & $E_g$ & $k_z(k_x\pm ik_y)$ & $E_u$ & $\pm i\hat{z}(k^{1+n}_x\pm ik^{1+n}_y);$\\
&   &   &   &$\pm ik_z(k^{n}_x\hat{x}\pm ik^{n}_y\hat{y});$\\
&   &   &   & $\pm ik_xk_yk_z(k^{n}_x\hat{y}\pm ik^{n}_y\hat{x})$,\\
&   &   &   &$n=0,2$\\\hline
\end{tabular}
\end{table}
In addition to superconducting channels, we also include particle-hole chanels and
set $\chi_{0l} = \sum_{\sigma} \langle c^{\dagger}_{i\sigma} c_{i+a \hat{l}\sigma}\rangle$
and $\chi_{1l} = \langle c^{\dagger}_{i\uparrow} c_{i+a \hat{l} \downarrow}\rangle$ with $l=x$ or $y$. When the hole concentration is $\delta$, the resulting mean-field Hamiltonian $H_{MF}$ can be written as
\begin{eqnarray}
 H_{MF}&=&\sum_{k\sigma}\xi_k c^{\dagger}_{k\sigma} c_{k\sigma}-\tilde{\alpha}_R
\sum_{k\sigma\sigma^{\prime}} c^{\dagger}_{k\sigma}\left(\hat{\sigma}\cdot\mathbf{g}_k
\right)_{\sigma\sigma^{\prime}} c_{k\sigma^{\prime}} \nonumber \\ &+&\Delta_s\sum_{k}\Gamma^{g}_k\left(c^{\dagger}_{k\uparrow} c^{\dagger}_{-k\downarrow}-c^{\dagger}_{k\downarrow} c^{\dagger}_{-k\uparrow}\right) \nonumber  \\
&+&\Delta_t\sum_{k}\Gamma^{u}_k c^{\dagger}_{k\uparrow} c^{\dagger}_{-k\uparrow}-\Gamma^{u*}_k c^{\dagger}_{k\downarrow} c^{\dagger}_{-k\downarrow},
\label{MF}
\end{eqnarray}
where the lattice constant $a$ is set to be 1, $\xi_k= -2 [ \tilde{t}(\cos k_x+\cos k_y)+2 \delta t^{\prime}\cos k_x\cos k_y+\delta t^{\prime\prime}(\cos 2k_x+\cos 2k_y) ]-\mu$ and $\mathbf{g}_{k}=(\sin k_y,-\sin k_x)$ with $\tilde{t}=\delta t+\frac{(J_1+J_3)}{2}\chi_0-\frac{J_2}{4}\chi_{1x}$ and $\tilde{\alpha}_R=\delta\alpha_R-(J_1+J_3)\chi_{1x}-\frac{J_2}{2}\chi_0$.
The pairing amplitudes $\Delta_{s}$ and $\Delta_{t}$ represent amplitudes for singlet and triplet pairing with corresponding gap functions $\Gamma^{g}_k$ and $\Gamma^{u}_k$ respectively. The allowed representations for gap functions are tabulated in the table \ref{table:C4}. In the absence of the Rashba spin-orbit interaction, it is known that $d$-wave in $B_{1g}$ representation and $s$-wave in $A_{1g}$ representation are two superconducting states with lowest energies. In the presence of the Rashba spin-orbit interaction, we find that $A_1$ and $B_1$ representations become the lowest two in energy due to mixing of the singlet and triplet pairings. For $s+p$ pairing symmetries in $A_1$ representation, we have $\Gamma^{g}_k=\cos k_x+\cos k_y$ and $\Gamma^{u}_k=\sin k_y+i\sin k_x$.  On the other hand,  for $d+p$ pairing symmetries in $B_1$ representation, we have $\Gamma^{g}_k=\cos k_x-\cos k_y$ and $\Gamma^{u}_k=\sin k_y-i\sin k_x$. The pairing amplitudes are related to the pairing fields by
$\Delta_s=-J_1\psi_{x}-\frac{J_2}{2}\Phi_x$ and 
$\Delta_p=\mp(\frac{J_2}{2}\psi_x-J_3\Phi_x)$ ( $-$ and $+$ refer to ​$A_1$ and $B_1$ representation respectively)
with the order parameters being given by $\psi_{l}\equiv\frac{1}{2}\langle c_{i+a \hat{l}\downarrow}c_{i\uparrow}-c_{i+a \hat{l} \uparrow}c_{i\downarrow} \rangle$ and $\Phi_x\equiv\frac{1}{2}\langle c_{i+a \hat{x}\uparrow}c_{i\uparrow}+c_{i+a \hat{x}\downarrow}c_{i\downarrow}\rangle$.
The mean-field parameters are solved using the following self-consistent equations
\begin{eqnarray}
 \psi_l&=&\frac{1}{N}\sum_{k}\langle c_{-k\downarrow}c_{k\uparrow}\rangle\cos k_l,\\
 \Phi_{x}&=&\frac{-i}{2N}\sum_{k}(\langle c_{-k\uparrow}c_{k\uparrow}\rangle+\langle c_{-k\downarrow}c_{k\downarrow}\rangle)\sin k_x,\\
 \chi_{0l}&=&\frac{1}{2N}\sum_{k\sigma}\langle c^{\dagger}_{k\sigma}c_{k\sigma}\rangle\cos k_l,\\
\chi_{1l}&=&\frac{i}{N}\sum_{k}\langle c^{\dagger}_{k\uparrow}c_{k\downarrow}\rangle \sin k_l\\
\delta&=&1-\frac{1}{N}\sum_{k\sigma}\langle c^{\dagger}_{k\sigma}c_{k\sigma}\rangle,
 \end{eqnarray}
where $\langle c_{-k\sigma}c_{k\sigma'} \rangle$ and  $\langle c^{\dagger}_{k\sigma}c_{k\sigma'}\rangle$ are computed with respect to the ground state of Eq.(\ref{MF}). Note that the triplet pairing can be cast into a gap function as $\hat{\Delta}(k)=i (\mathbf{d}_k\cdot\sigma) \sigma_y$ with $\mathbf{d}_k$ being the $d$-vector.
In the $B_1$ representation, the gap function $\mathbf{d}_k$ vector is not aligned with the internal magnetic field $\mathbf{g}_k$. We find that the non-aligned $d$-vector lower the ground state energy in consistent with previous studies\cite{Sigrist2}. When $J_3$ dominates, we find that the spin-triplet pairing wins over so that $A_1$ representation dominates. Hence as $J_3$ increases, superconducting phases will go through a transition from $d+p$ wave ($B_1$) to $s+p$  wave ($A_1$). In the $A_1$ representation,  $\mathbf{d}_k$ vector is parallel to $\mathbf{g}_k$ vector, the ground state energy is further lowered down  in comparison to the $B_1$ representation.
\begin{figure}[htp]
\includegraphics[width=7.0cm]{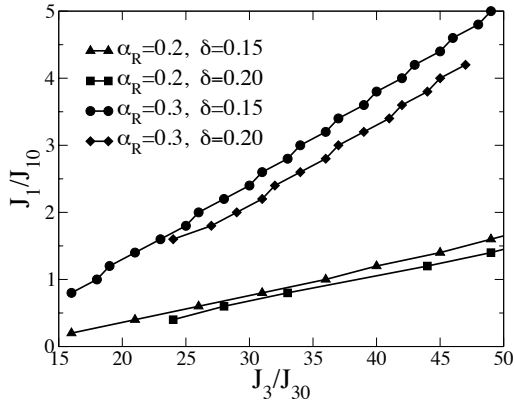}\\
\caption{Phase boundary between  $d+p$ wave (small $J_3$) to $s+p$  wave (large $J_3$). Here $J_{10}=\frac{4t^2}{U}$ and $J_{30}=-\frac{\alpha^2_R}{U}$. Unless $J_3$ is very large, for typical strength of Rashba interaction, the pairing symmetry is $d+p$ wave. }
\label{Fig2}
\end{figure}

\begin{figure}[tp]
\hspace{-0.2cm}\includegraphics[width=8.5cm]{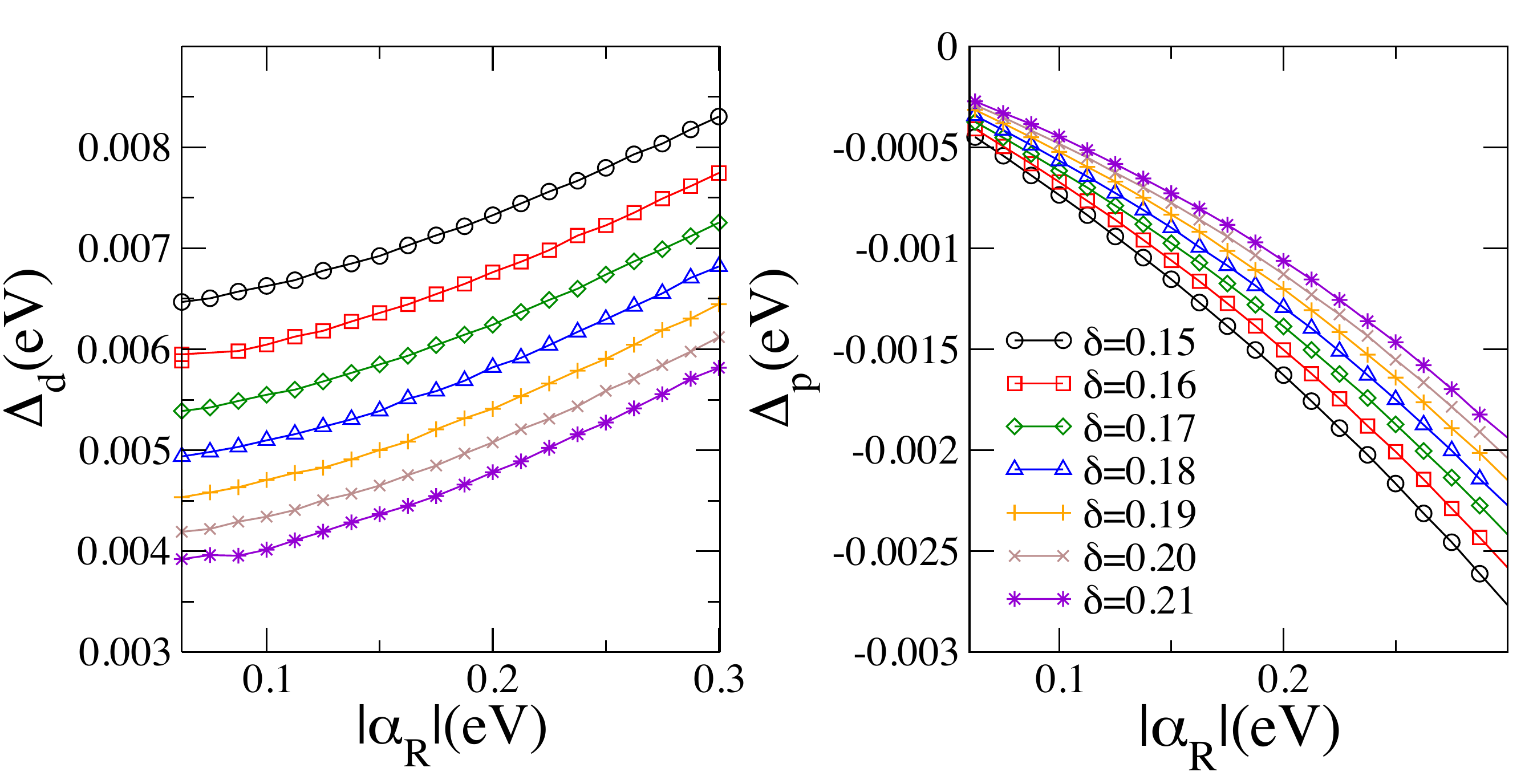}\\
\caption{Mean-field pairing amplitudes for $d$-wave and $p$-wave versus Rashba spin-orbit coupling strength. Here parameters are $t=0.45, t^{\prime}=-0.1575, t^{\prime\prime}=0.0788$ in units of $eV$.}
\label{Fig3}
\end{figure}

\begin{figure}[htp]
\begin{center}
\includegraphics[width=7.0cm]{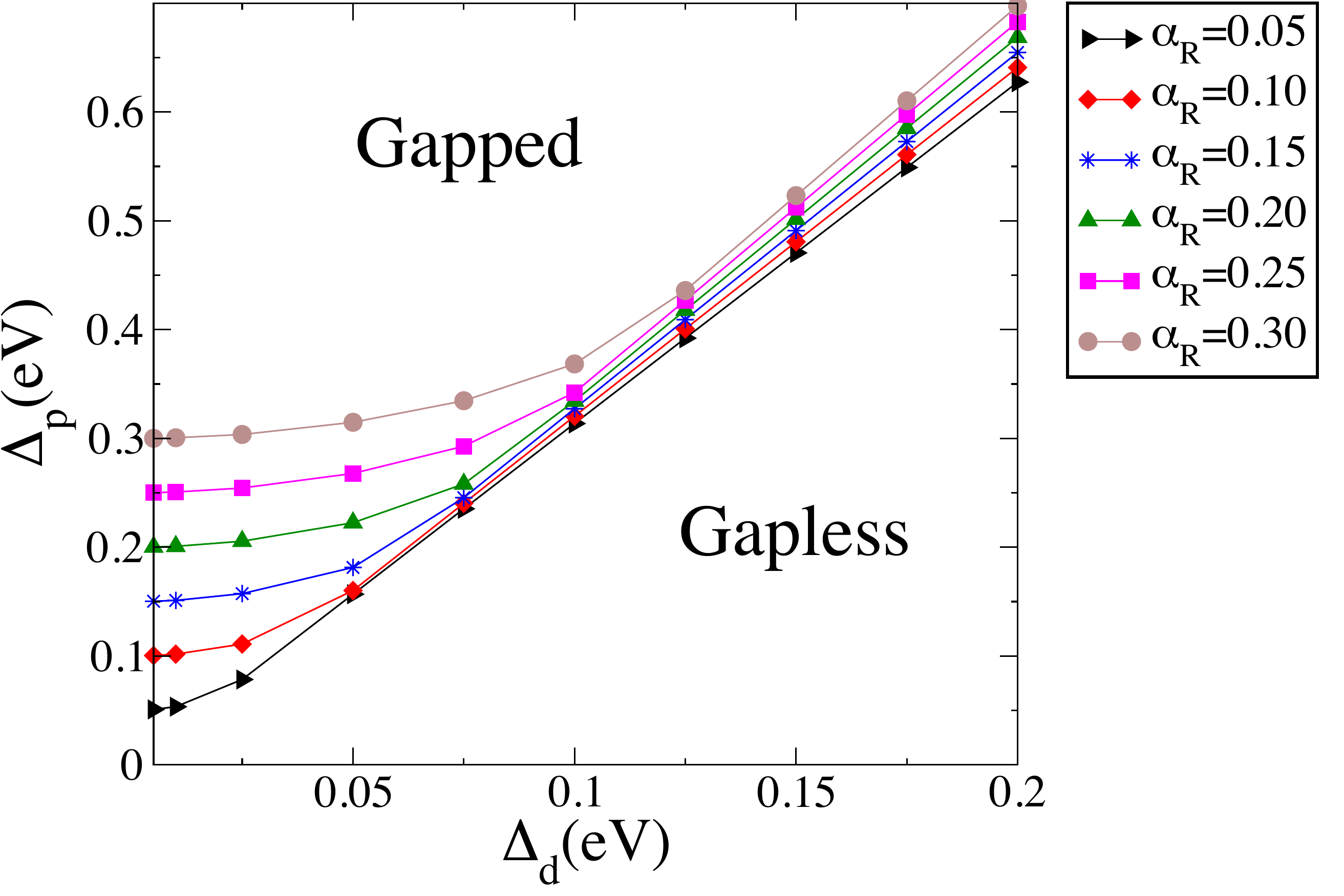}
\caption{Typical boundaries between gapped and gapless superconducting phases with $\alpha_R =0.05-0.3$ ($a$ is the lattice constant). Here $\mu$ is fixed at $-0.9$. It is seen that for realistic parameters $\Delta_d \ge \Delta_p$, the superconducting phase is gapless.}
\label{Fig4}
\end{center}
\end{figure}

Fig.~\ref{Fig2} shows the phase boundary between $d+p$ wave ($B_1$) to $s+p$  wave ($A_1$) for different dopings at a fixed $J_2$. The upper left corner of each is $d+p$ wave while the lower right corner is $s+p$ wave.  It is clear that in the strong coupling limit, the pairing symmetry is $s+p$ wave in $A_1$ only when  $J_3$ is at least $10$ times $J_{30}$ for $J_1=\frac{4t^2}{U}$. For For typical strength of Rashba interaction, the pairing symmetry is $d+p$ wave. Fig.~\ref{Fig3} examines how $d$-wave and $p$-wave amplitudes change as the Rashba interaction $\alpha_R$ increases. Due to that the $J_3$ term provides spin-triplet pairing channel and $J_3\propto\alpha_R^2$, the $p$-wave amplitude increases quadratically, $\Delta_p\propto\alpha_R^2$.  However, for typical strength of $\alpha_R$, the amplitude of $p$-wave is still one order of magnitude less that that of $d$-wave. On the other hand, the amplitude of $d$-wave also increases linearly with $\alpha_R$ due to the coupling of $d$-wave to $p$-wave via the $J_2$ term,
which is proportional to $\alpha_R$. As a result, $d$-wave component is always larger than that of $p$-wave component. Hence for realistic parameter regimes, the resulting superconducting state is still dominated by $d$-wave component and the ground state is still in gapless phase. As an example, we show the boundary between gapped phase and gapless phase for the largest reported value of $\alpha_R$ in Fig.~\ref{Fig4}. It is seen that for realistic parameters $\Delta_d \ge \Delta_p$, the superconducting phase is gapless.

\section{Topological gapless phases and edge states}
\label{gapless}

In this section, we investigate the gapless regime of $d+p$ wave. This is the regime that typical strength of Rashba interaction lies in. It is shown that nodal points split and
the resulting superconducting state always supports edge states for any interfaces that are not exactly aligned in $(100)$ or $(010)$ directions.
 
\subsection{Bulk nodal points and their transitions in structure}
We start from the general bulk Hamiltonian with $d+p$ mixed pairing symmetry 
and Rashba interaction in 
the Nambu basis $H=\sum_{\mathbf{k}}\hat{\Psi}^{\dagger}_{\mathbf{k}} h (\mathbf{k})\hat{\Psi}_{\mathbf{k}}$, 
where $\hat{\Psi}_{\mathbf{k}} =
(c_{\mathbf{k}\uparrow},c_{\mathbf{k}\downarrow},c^{\dagger}_{-\mathbf{k}\uparrow},c^{\dagger}_{-\mathbf{k}\downarrow})^T$. The $4\times4 $ matrix form of $h(\mathbf{k})$ is given by
\begin{eqnarray}h(\mathbf{k})=\left (
\begin{array}{cccc}
 \xi_k& \alpha_R\Lambda_k & \Delta_p\Lambda^{*}_k&\psi_k\\
 \alpha_R\Lambda^*_k & \xi_k & -\psi_k &-\Delta_p\Lambda_k   \\
 \Delta_p\Lambda_k & -\psi_k&-\xi_k&\alpha_R\Lambda^*_k \\
 \psi_k&-\Delta_p\Lambda^{*}_k &\alpha_R\Lambda_k&-\xi_k
\end{array}\right ),
\end{eqnarray} 
where the lattice constant $a$ is set to be 1, $\Lambda_k=\sin k_y+i\sin k_x$, the kinetic energy is $\xi_k=-2t(\cos k_x+\cos k_y)-4t^{\prime}\cos k_x\cos k_y-2t^{\prime\prime}(\cos 2k_x+\cos 2k_y)-\mu$, and $\psi_k$ is the singlet-pairing $d$-wave gap function.  The gap function can be generally combined as $\hat{\Delta}(k)=i(\psi_k+\mathbf{d}_k\cdot\sigma)\sigma_y$ with $\psi_k=-\Delta_d(\cos k_x-\cos k_y)$ and $\mathbf{d}_k=-\Delta_p(\sin k_y,\sin k_x,0)$. The energy spectra $E_k$ can be solved and are given by
\begin{eqnarray}\label{Ek}
 E_k=\pm\sqrt{\xi^2_k+\psi^2_k+\mathbf{g}^2_k+\mathbf{d}^2_k\pm2\sqrt{\mathbf{A}^2_k+
\mathbf{B}^2_k}}, \label{spectra}
\end{eqnarray}
where $\mathbf{g}_k=\alpha_R(\sin k_y,-\sin k_x)$, $\mathbf{A}_k\equiv\xi_k\mathbf{g}_k+\psi_k\mathbf{d}_k$ and $\mathbf{B}_k\equiv\mathbf{g}_k\times\mathbf{d}_k$.

\begin{figure}[htp]
\begin{center}
\includegraphics[width=7.5cm]{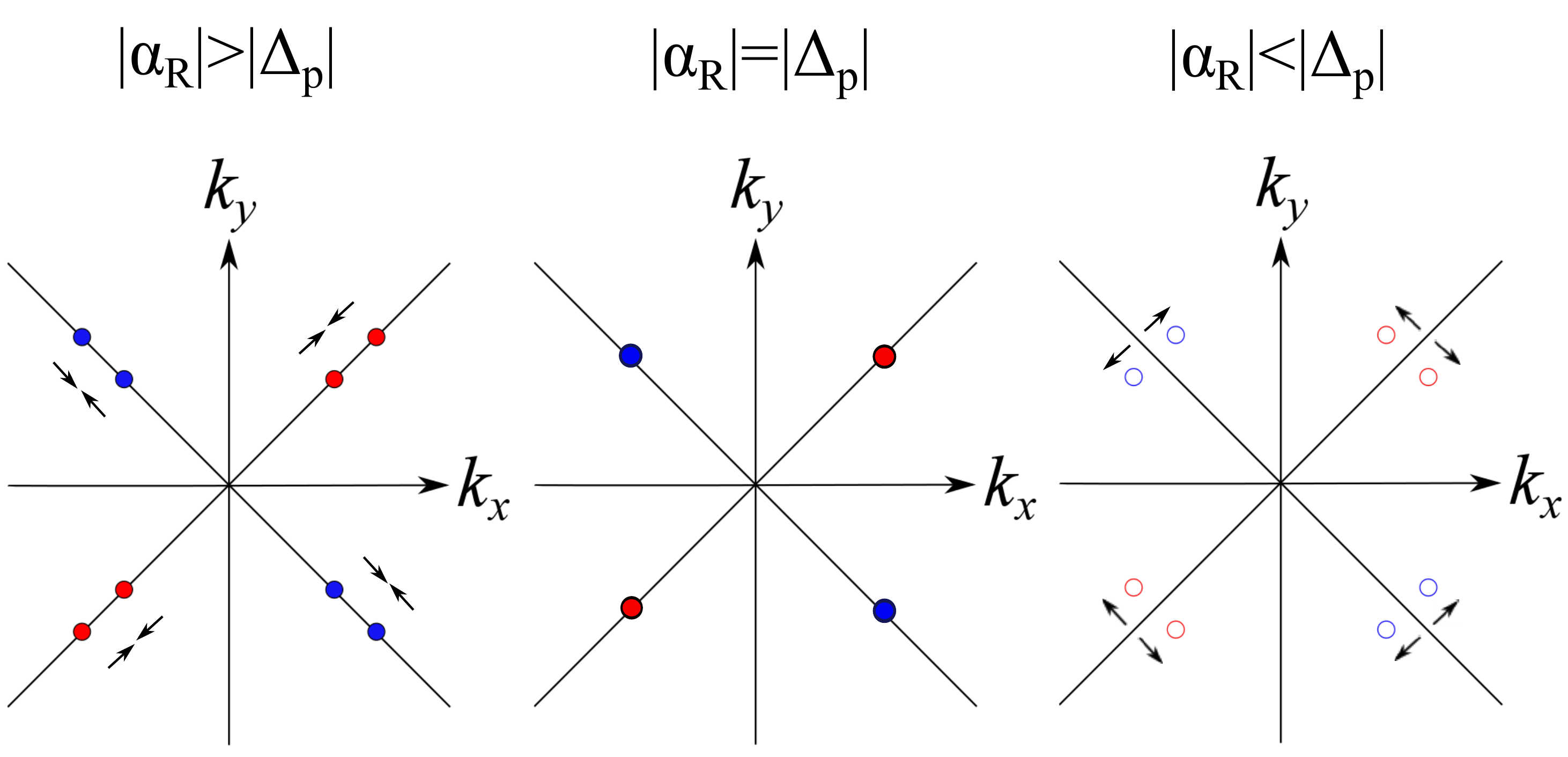}
\caption{Nodal point structure in the presence of Rashba spin-orbit interaction. Here solid circles are nodal points for $|\alpha_R|>|\Delta_p|$, open circles are nodal points for $|\alpha_R|<|\Delta_p|$ and when $|\alpha_R|=|\Delta_p|$, nodal points in each quadrant recombine.}\label{Fig5}
\end{center}
\end{figure}

In the absence of the Rashba spin-orbit interaction and the $p$-wave, 
the Hamiltonian $h_0(\mathbf{k})$ supports 4 Dirac points determined  $\xi_k=0$ and $\psi_k=0$.  The 
$\mathbf{k} \cdot \mathbf{p}$ expansion of $h (\mathbf{k})$ about one Dirac point $\mathbf{k}_D$ can be generally written as
\begin{eqnarray}
 h_0(\mathbf{k})= \sum^3_{i=1} v_{F_i} \gamma_i q_i,
\end{eqnarray}
where $\mathbf{q} = \mathbf{k}-\mathbf{k}_D$ and coordinate axes are along diagonal axes with $q_1=q_x+q_y$ and $q_2=q_x-q_y$. For $\mathbf{k}_D$ in the first quadrant, one finds that $\mathbf{k}_D=(k_{D_x},k_{D_y},0)=(k_{D_x},k_{D_x},0)$. In the diagonal coordinates, we have
$v_F=(2t \sin k_{D_x}+(2t'+4t'')\sin 2 k_{D_x}, -\Delta_d \sin k_{D_x}, 0)$,  $\gamma_1=s_z$, and $\gamma_2=s_y \sigma_y$. Here $\mathbf{s}$ is the pseudospin vector in the particle-hole space. 
In analogy to the characterization of 3D Weyl semimetals\cite{Burkov} and Weyl fermions in ferromagnetic superconductors\cite{Tewari}, 
$\gamma_1$ and $\gamma_2$ can be extended to be
a complete set of $\gamma$ matrices in SU(4) representation $\{ I, \gamma_i, \gamma_{ij}\}$ with  $\{ \gamma_1,\gamma_2,\gamma_3,\gamma_4,\gamma_5 \}=\{ s_z,s_y \sigma_y, s_y \sigma_x, s_x, s_y\sigma_z \}$. 
If we redefine operations of time-reversal and inversion symmetries as operators with respect to $\mathbf{q}$ for
small $q$,  the Rashba spin-orbit interaction and $p$-wave order parameter break time-reversal symmetry (operator represented by $i\sigma_y K$) and inversion symmetry (operator represented by $s_x$). In this case, perturbations due to broken time-reversal and inversion symmetries can be generally expressed in terms of $\gamma_{ij}$ matrices defined by $\gamma_{ij}=-i/2[\gamma_i,\gamma_j]$ \cite{Burkov}. We find that
\begin{eqnarray}
 h(\mathbf{k})= h_0(\mathbf{k}) +\mathbf{u} \cdot \mathbf{b} + \mathbf{v} \cdot \mathbf{b}', \label{perturb}
\end{eqnarray}
where $\mathbf{b} = (\gamma_{14},\gamma_{24},\gamma_{34})$, $\mathbf{b}'= (\gamma_{15},\gamma_{25},\gamma_{35})$, $\mathbf{u}=-\sin k_{D_x}(-\Delta_p, \alpha_R,0 )$, and $\mathbf{v}=\sin k_{D_x}(-\Delta_p, \alpha_R,0 )$. The corresponding energy spectrum $E_k$ of Eq.(\ref{perturb}) becomes $E_q$ and is given by
\begin{eqnarray}
 E^{\pm}_q = & & \pm \sqrt{p^2_1+p^2_2+2(\alpha^2_R+\Delta^2_p)\sin^2k_{D_x} \pm} \nonumber\\
&& \overline{2\sqrt{2(\alpha^2_R p^2_1+\Delta^2_p p^2_2)\sin^2k_{D_x}+4\alpha^2_R \Delta^2_p \sin^4k_{D_x}}}, \nonumber \\
\end{eqnarray}
where we have defined $p_1= v_{F_1} q_1$ and $p_2= v_{F_2} q_2$.
Clearly, when $q_1=q_2=0$, $E_q$ no longer vanishes. Hence the original nodal points are lifted. Similar to the 3D Weyl fermions, the main effect of the Rashba interaction and the $p$-wave is to split each Dirac point into two Weyl nodal points in 2D. Specifically, $h(\mathbf{k})$ obtained in Eq.(\ref{perturb}) corresponds to the case $\mathbf{u} \parallel \mathbf{v}$ for 3D Weyl semimetal\cite{Burkov}. We find that for $|\alpha_R| > |\Delta_p |$, $E_q$ vanishes at $p_1 = \pm \sqrt{2(\alpha^2_R-\Delta^2_p)\sin^2k_{D_x}}$ and $p_2=0$; while for 
$|\alpha_R| < |\Delta_p |$, $E_q$ vanishes at $p_1 =0$ and $p_2= \pm \sqrt{2(\Delta^2_p -\alpha^2_R)\sin^2k_{D_x}}$.

The transformation of nodal point structure is illustrated in Fig.~\ref{Fig5}. For a fixed $\alpha_R$, as $\Delta_p$ increases, a transition of nodal point configuration occurs at $|\alpha_R|=|\Delta_p|$ before the ground state becomes a full gapped TRI superconducting state. The transition of nodal point structure results from competition between $d$-wave and $p$-wave pairing symmetries. For small $p$-wave amplitudes when $|\alpha_R|>|\Delta_p|$, the superconducting state is dominated by the $d$-wave symmetry. Hence nodal points are determined by nodal lines of $d$-wave order parameter, which are along axes of $k_y = \pm k_x$. The intersections of splitting Fermi surfaces by $\alpha_R$ and nodal lines of $d$-wave order parameter determines nodal points along $k_y = \pm k_x$. For large $p$-wave amplitudes when $|\alpha_R|<|\Delta_p|$, nodal points are determined by nodal lines of $p$-wave order parameter, which are $k_x =0$ or $k_y=0$. As a result, nodal points are along the $q_1$($q_2$) axis in perpendicular to the axes of $k_y = k_x$ ($k_y=-k_x$).

\subsection{Bulk nodes and edge states}

\begin{figure}[htp]
\begin{center}
\includegraphics[width=4.2cm]{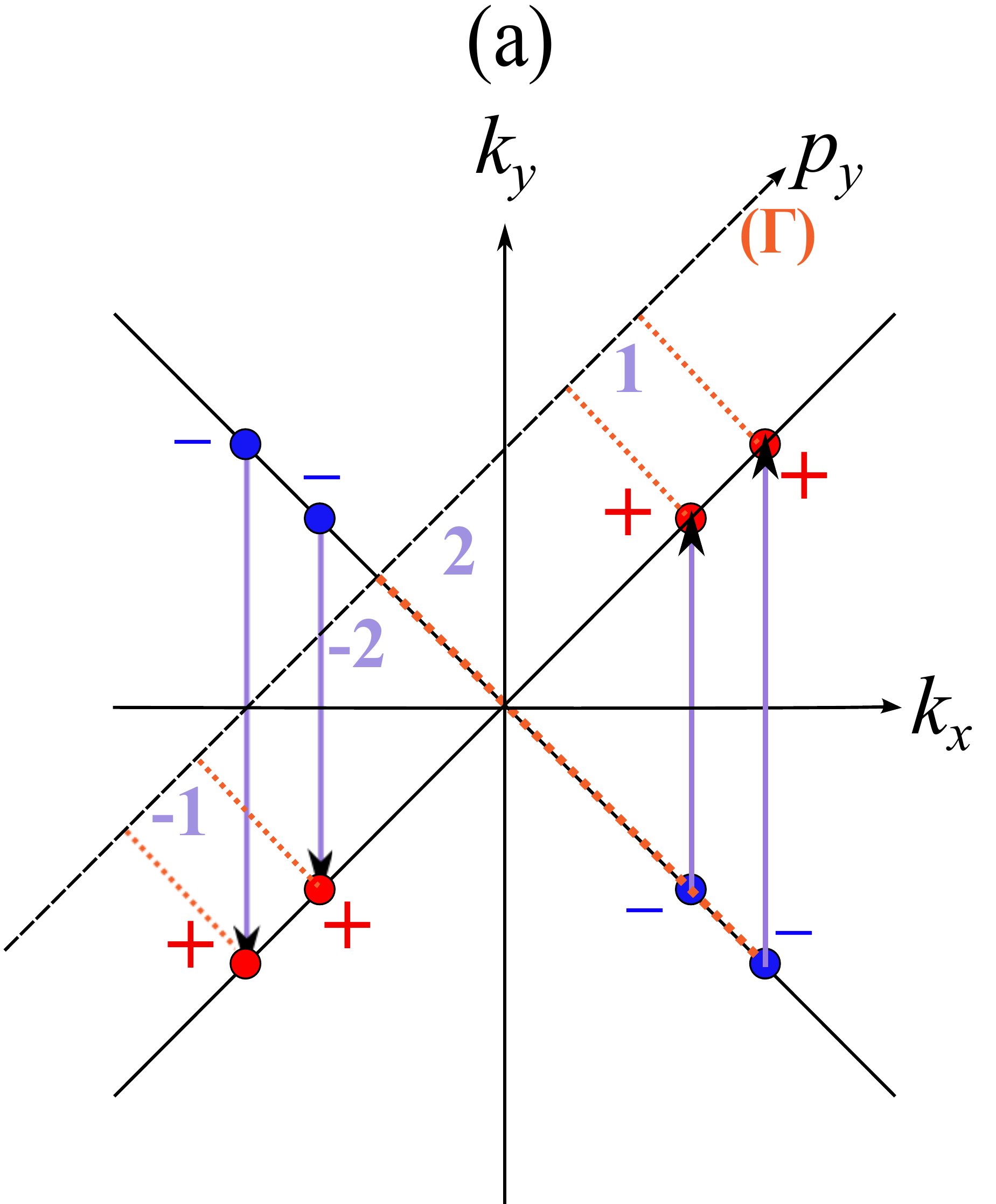}
\includegraphics[width=4.2cm]{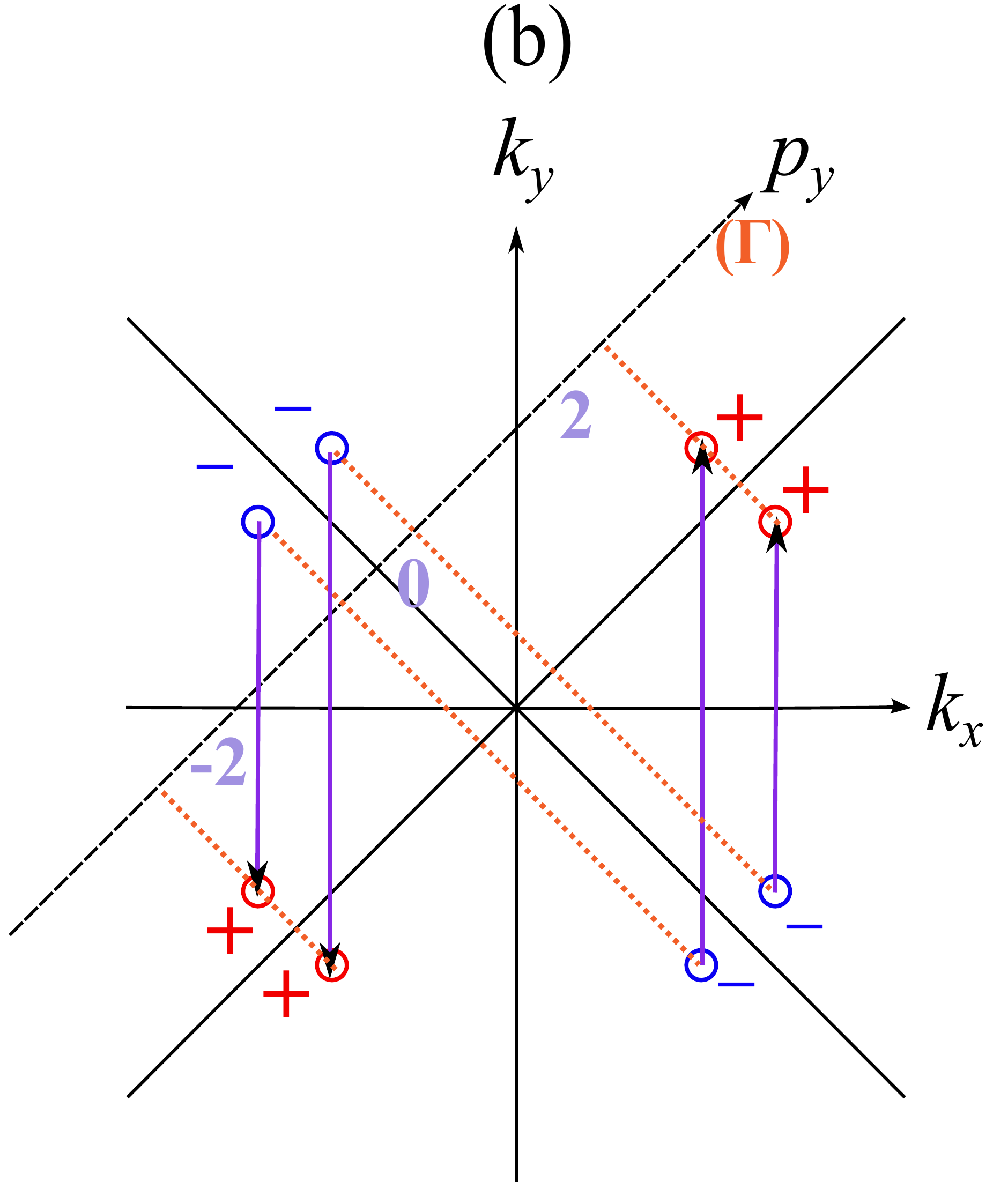}
\caption{Projection of nodal points with winding numbers labelled by $\pm$ for (a) $|\alpha_R|>|\Delta_p|$
and (b) $|\alpha_R|<|\Delta_p|$. Here each nodal point is associated with a branch cut in $-k_y$ direction with phase jump being $\pm 2 \pi$. Due to the cancellation of phase jumps, only lines that connect $+$ and $-$ are branch cuts with non-vanishing phase jumps. The projection of each line connecting $+$ and $-$ onto the edge $\Gamma$ axis gives rise to one zero-energy flat band at the projected range of wavevector $p_y$. Absolute values of numbers ($\pm 2$ or $\pm 1$) shown in the edge $\Gamma$ axis are numbers of flat-bands in the corresponding $p_y$ range.  It is seen that for realistic Rashba strength, $|\alpha_R|>|\Delta_p|$, one always gets single zero-energy mode for certain wavevector $p_y$ along the edge $\Gamma$.}
\label{Fig6}
\end{center}
\end{figure}

The nodal points revealed in the last subsection carry winding numbers.  The associated winding numbers with nodes labeled by $\pm$ are shown in Fig.~\ref{Fig6}. When positions of nodal points change, the distribution of winding numbers also changes. It is known that positions of nodal points determine edge states. According to Refs.\onlinecite{DHLee} and \onlinecite{Volovik}, for 3D bulk states, as long as the projection of opposite-winding-number nodal manifolds does not completely overlap in the boundary Brillouin zone, there will be gapless surface bound states. For 2D bulk states, the gapless surface bound states become zero-energy edge states. Therefore, for a given edge, the projection of nodal point to the edge determines zero-energy edge states.

Fig.~\ref{Fig6} illustrates the projection of nodal points to edges. For each nodal point, one associates a branch cut with phase jump being $\pm 2 \pi$. In Fig.~\ref{Fig6}, branch cuts are chosen in $-k_y$ direction. Due to the cancellation of phase jumps,
only lines that connect $+$ and $-$ are branch cuts with non-vanishing phase jumps. Here the phase jumps for the line connecting  $+$ to $-$ and that for $-$ to $+$ are opposite in sign. By projecting branch cuts onto the momentum line (denoted by $p_y$) that represents the edge,
it is clear that for $|\alpha_R|>|\Delta_p|$ shown in Fig.~\ref{Fig6}a, the nodal manifolds for $\pm$ nodes do not overlap completely except when orientations of edges are exactly along $(100)$ or $(010)$ directions. Therefore, for $p_y$ in the non-overlapped region, there always exists single zero-energy mode along the edge. The single zero-energy mode is a Majorana mode. More specifically, the projection of each connection of $+$ and $-$ to a given edge $\Gamma$ gives rise to one zero-energy mode for each wavevector $p_y$ along $\Gamma$ that lies in the non-overlapped region. It thus results in a flat band at zero-energy\cite{Sato}. Therefore, in the overlapped region, there will be two flat-bands; while in the non-overlapped region, there is only one branch flat-band which is a Majorana flat-band. As a result, we find that except when the edge is exactly along $(100)$ or $(010)$ directions, in general, there always exist Majorana modes for any edges. Since real interfaces always possess facets in different orientations, existence of Majorana modes appear to be a robust feature.

As the amplitude of $p$-wave increases until $|\Delta_p|>|\alpha_R|$, nodal points switch to be in the $k_2$ direction. As illustrated in Fig.~\ref{Fig6}b, the nodal manifolds for $\pm$ nodes do not overlap completely except when orientations of edges are exactly along $(110)$ or $(\bar{1}10)$ directions.  Therefore, except when the edge is exactly along $(110)$ or $(\bar{1}10)$ directions, in general, there always exists Majorana modes for any edges.  Note that in the case when $|\Delta_p|>|\alpha_R|$, there is a region near $p_y=0$ for $(110)$ edge in which as we shall show in below,  the connection of nodal points with the same winding number leads to dispersive Majorana modes. 


\subsection{Edge states along $(110)$ edge}
In this subsection, we shall verify connections of edge states to the nodal structure for the particular orientation of edge $(110)$.
The $(110)$ edge is known to be the most important edge that exhibits zero-bias peak in the tunneling spectrum due to the existence of Andreev bound states at zero energy\cite{Mou1}. It is therefore important to  examine how edge states change when the Rashba spin-orbital interaction is included. 
We shall start by partial Fourier transforming the Hamiltonian along the $(110)$ interface direction.
Let $p_y$ be the Fourier wavevector along $(110)$ direction. The Hamiltonian can be expressed as
$H= \sum_{p_y}\hat{\Psi}^{\dagger}(p_y) H_{1D}(p_y) \hat{\Psi}(p_y)$ and is characterized by an one-dimensional Hamiltonian $H_{1D}(p_y)$ for a given $p_y$, which is given by
\begin{eqnarray}
H_{1D}(p_y) &=&\left (
\begin{array}{cc}
 \hat{h}(p_y) & \hat{\Delta}(p_y) \\
 \hat{\Delta}^{\dagger}(p_y) & -\hat{h}^{T}(-p_y)  \\
\end{array}\right ),  
\label{H1D}
\end{eqnarray}
where the hopping and pairing matrices $\hat{h}(p_y)$ and $\hat{\Delta}(p_y)$ are given by
\begin{eqnarray}
\hat{h}(p_y)=\left(\begin{array}{cc}
 \hat{h}^{\uparrow\uparrow}_{p_y} & \hat{h}^{\uparrow\downarrow}_{p_y} \\
 \hat{h}^{\downarrow\uparrow}_{p_y} & \hat{h}^{\downarrow\downarrow}_{p_y}  
\end{array}\right),\hat{\Delta}(p_y)=\left (
\begin{array}{cc}
 \hat{\Delta}^{\uparrow\uparrow}_{p_y} & \hat{\Delta}^{\uparrow\downarrow}_{p_y} \\
 \hat{\Delta}^{\downarrow\uparrow}_{p_y} & \hat{\Delta}^{\downarrow\downarrow}_{p_y} 
\end{array}\right ).
\end{eqnarray}
The block matrices  $\hat{h}(p_y)$ are expressed as 
\begin{eqnarray}
\nonumber\hat{h}^{\uparrow\uparrow}_{p_y}=\left (
\begin{array}{ccccc}
 t_{\perp} & t_{\parallel} & t^{\prime}_{\parallel} &0 &\dots \\
 t_{\parallel} &  t_{\perp} &  t_{\parallel} & t^{\prime}_{\parallel} &\dots \\
t^{\prime}_{\parallel} & t_{\parallel} & t_{\perp} &  t_{\parallel} &t^{\prime}_{\parallel}\\
0 & t^{\prime}_{\parallel} & t_{\parallel} & t_{\perp} & t_{\parallel}\\
\vdots & 0 &t^{\prime}_{\parallel} & t_{\parallel} & \ddots
\end{array}\right ),\hat{h}^{\uparrow\downarrow}_{p_y}=\left (
\begin{array}{cccc}
 0 & t_{1} & 0 &\dots \\
 t_{2} & 0 &  t_{1} & 0  \\
0 & t_{2} & 0&  t_{1} \\
\vdots & 0 &t_{2} & \ddots
\end{array}\right ),
\end{eqnarray}
with $\hat{h}^{\downarrow\downarrow}_{p_y}=\hat{h}^{\uparrow\uparrow}_{p_y}$ and $\hat{h}_{p_y}^{\downarrow\uparrow}=(\hat{h}^{\uparrow\downarrow}_{p_y})^{\dagger}$. The pairing matrices satisfy $\hat{\Delta}_{\uparrow\downarrow}(p_y)=-\hat{\Delta}_{\downarrow\uparrow}(p_y)$ and $\hat{\Delta}^{\downarrow\downarrow}_{p_y}=-(\hat{\Delta}^{\uparrow\uparrow}_{p_y})^{\dagger}$ with
\begin{eqnarray}
\nonumber\hat{\Delta}^{\uparrow\downarrow}_{p_y}=\left (
\begin{array}{cccc}
 0& \Delta_1 & 0 &\dots\\
 \Delta_2 &  0 &  \Delta_1 & 0 \\
0 & \Delta_2 & 0 &  \Delta_1 \\
\vdots & 0 &\Delta_2&\ddots
\end{array}\right ),
\hat{\Delta}^{\uparrow\uparrow}_{p_y}=\left (
\begin{array}{cccc}
 0 & \Delta_3 & 0 &\dots \\
 \Delta_4 &  0 &  \Delta_3 & 0   \\
0 & \Delta_4 & 0&  \Delta_3\\
\vdots & 0 & \Delta_4&\ddots
 \end{array}\right ).\end{eqnarray}
Here elements in matrices are given by $t_{\perp}=-2t^{\prime}\cos(p_y\delta^{\prime}_{2y})-\mu$, $t_{\parallel}=-2t\cos (p_y\delta_{1y})$, $t^{\prime}_{\parallel}=-t^{\prime}-2t^{\prime\prime}\cos (p_y\delta^{\prime\prime}_{1y})$, $t_1=\frac{\alpha_R}{2}(-ie^{ip_y\delta_{1y}}+e^{-ip_y\delta_{1y}})$, $t_2=\frac{\alpha_R}{2}(ie^{-ip_y\delta_{1y}}-e^{ip_y\delta_{1y}})$, $\Delta_1=-\Delta_2=i\Delta_d\sin (p_y\delta_{1y})$, $\Delta_3=\frac{\Delta_p}{2}(-ie^{ip_y\delta_{1y}}-e^{-ip_y\delta_{1y}})$, $\Delta_4=\frac{\Delta_p}{2}(ie^{-ip_y\delta_{1y}}+e^{ip_y\delta_{1y}})$ with $\delta_{1y}=1/2$ and $\delta^{\prime}_{2y} =\delta^{\prime\prime}_{1y}=1$. 
\begin{figure}[htp]
\center
\includegraphics[width=8.5cm]{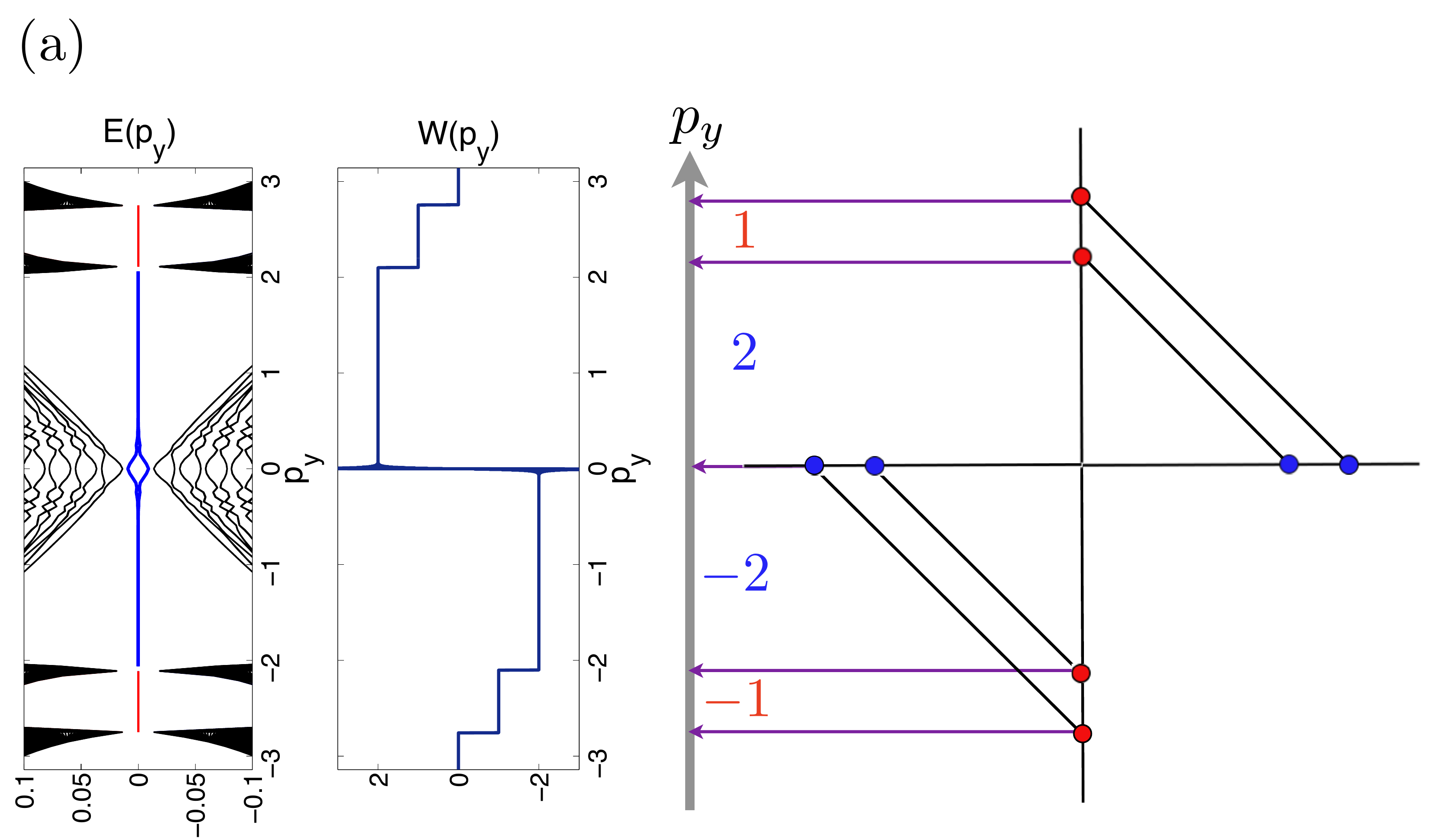}
\center
\includegraphics[width=8.5cm]{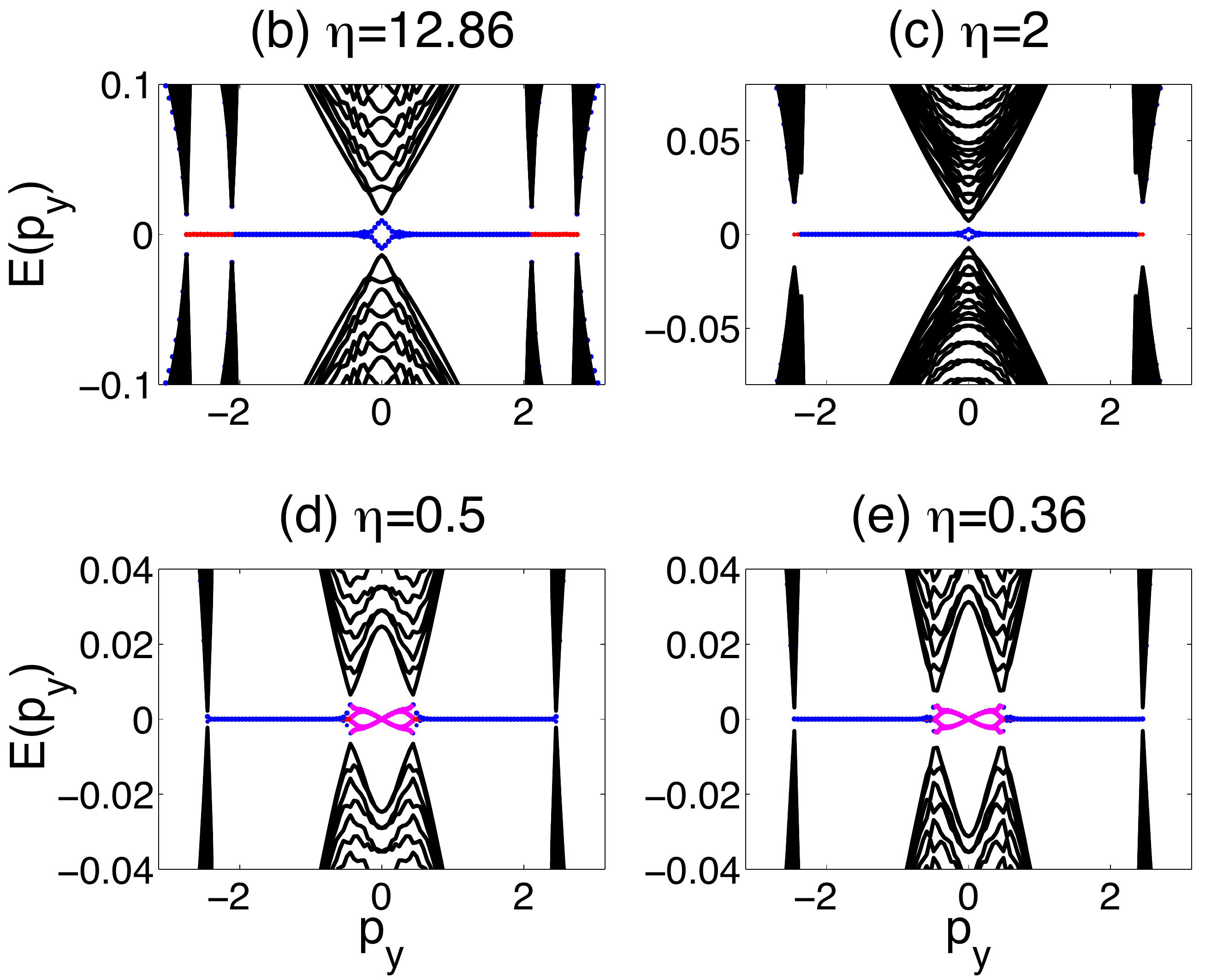}
\caption{(a) Comparison and consistent check of analysis by projection and analysis by winding number
for $(110)$ edge.
(b)-(e) Evolution of quasi-particle spectrum and edge states for $(110)$ edge of cuprates in the gapless phase when $\alpha_R$ changes.  Here $\eta = \alpha_R/\Delta_p$. The mean-field solution corresponds to $\eta=12.86$ with $\delta=0.16$ in (b).  Edge states are labelled as blue and red solid lines with red lines representing Majorana edge states. It is clear that Majorana flat-bands at zero energy exist only for physical accessible region $\alpha_R > \Delta_p$; while the dispersive Majorana modes exist near $p_y=0$ only for  $\alpha_R < \Delta_p$.}
\label{Fig7}
\end{figure}

\begin{figure}[htp]
\center
\includegraphics[width=8cm]{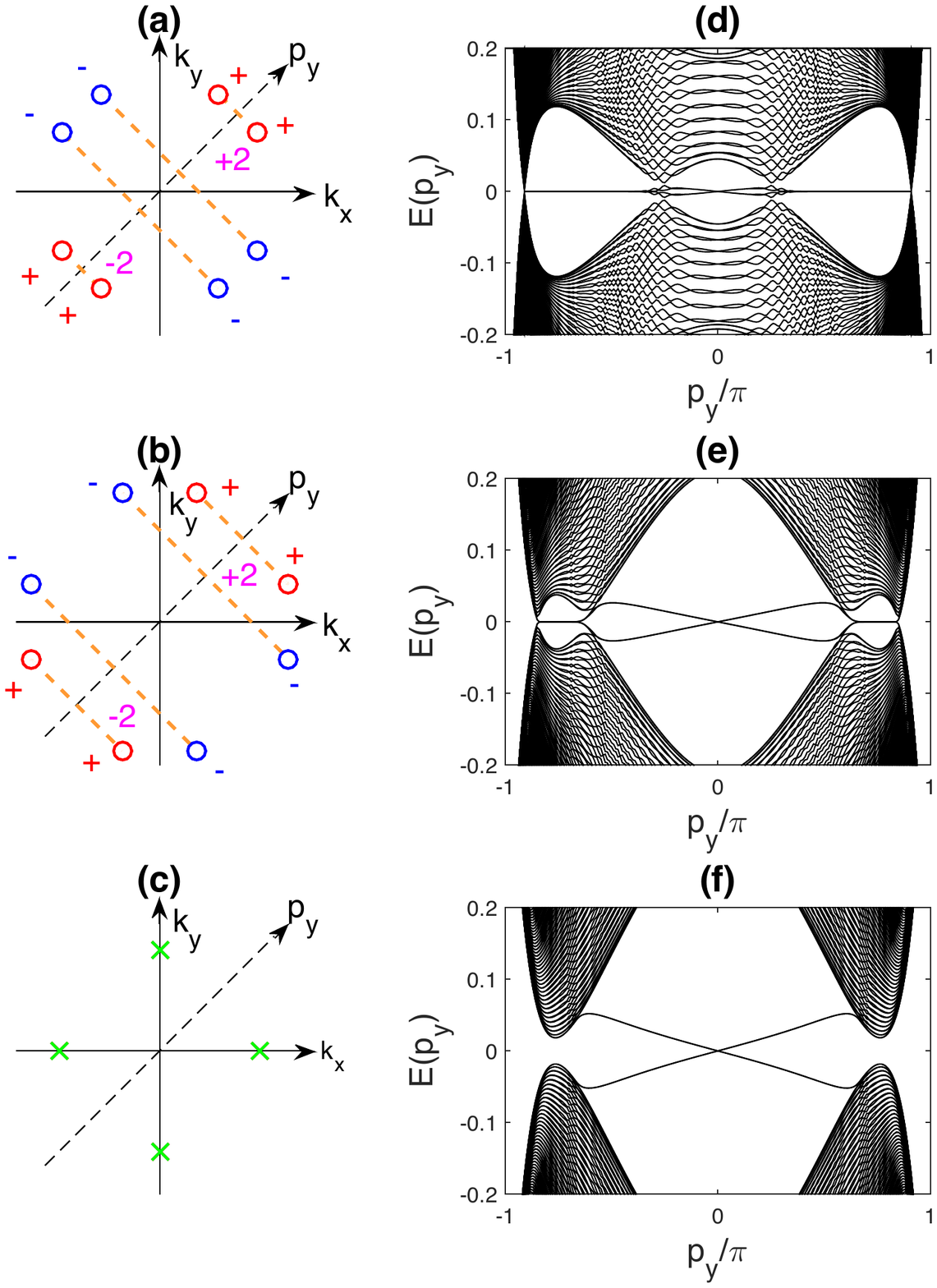}
\caption{Illustration of the nature of dispersive edge states for $(110)$ edge in the gapless phase. Here $t=1, t^{\prime}=t^{\prime\prime}=0, \mu=0.5$,
and  $\Delta_p > \alpha_R $ with $\Delta_d=0.1$ and $\alpha_R=0.05$ being holding fixed.  
As shown in (a)-(c), as $\Delta_P$ increases ($\Delta_p =0.08$, $0.2$ and $0.28$), $\pm$ nodal points from different quadrants move toward each other and annihilate. Correspondingly, as shown in (d)-(f), the region with dispersive edge modes around $p_y=0$ expands and the region (labelled by $\pm 2$) with fermionic Andreev bound states shrinks. It is clear that dispersive edge modes near $p_y=0$ evolve into helical Majorana modes of the fully gapped $p$-wave superconductors.}
\label{Fig8}
\end{figure}
We shall focus on the case when the superconducting side is semi-infinite. For a given $p_y$, the system 
becomes one-dimensional and its topology can be characterized by the winding number\cite{Sato}. Here we extend the winding number calculation\cite{Sato} to any edge that makes an angle $\theta$ clock-wisely with respect
to $(100)$ direction.  The winding number $W$ is then given by
\begin{eqnarray}
W(p_{\parallel},\theta)&=&\frac{1}{2\pi}Im\int^{\pi}_{-\pi} dp_{\perp}\partial_{p_{\perp}}\ln \det \hat{q}(p_{\parallel},p_{\perp},\theta), \nonumber \\
&=&\nonumber \frac{1}{2\pi}\int^{\pi}_{-\pi} dp_{\perp}\epsilon^{ab}m_{a}(p_{\parallel},p_{\perp},\theta)\partial_{p_{\perp}}m_{b}(p_{\parallel},p_{\perp},\theta). \nonumber \\
\end{eqnarray}  
Here $p_{\parallel}$ is the wavevector along the edge, $p_{\perp}$ is the wavevector perpendicular to the edge, and the matrix $\hat{q}$ is given by $\hat{q}=i(\xi(\mathbf{p},\theta)-i\psi(\mathbf{p},\theta))\sigma_y+i(\vec{g}(\mathbf{p},\theta)-i\vec{d}(\mathbf{p},\theta))\cdot\sigma\sigma_y$.
$m_1(\mathbf{p},\theta)$ and $m_2(\mathbf{p},\theta)$ are the real and imaginary parts of the phase factor of $\det \hat{q}(\mathbf{p},\theta)$ respectively. 

For a give $p_y$, if number of sites in perpendicular to $(110)$ edge is $N$, $H_{1D}(p_y)$ is a $4N\times 4N$ matrix and can be exactly diagonalized. For large $N$, one obtains numerical solutions for semi-infinite superconducting state. In Fig.~\ref{Fig7}(a), we show how the quasi-particle spectrum for the $(110)$ edge changes as the Rashba spin-orbit interaction $\alpha_R$ varies.  Starting from the mean-field solution with $\eta \equiv \alpha_R/\Delta_p =12.86$ shown in  Fig.~\ref{Fig7}(a), there are two Majorana flat-bands with winding number $|W|=1$ at $2 \lesssim p_y \lesssim 2.6$. The decreasing of $\alpha_R$ leads to the reduction of the Majorana flat-band regimes as shown Fig.~\ref{Fig7}(b). In particular, as indicated in Fig.~\ref{Fig7}(c) and (d), Majorana flat band disappears for $\eta \le 1$ (i.e. $\alpha_R \le \Delta_p$). Instead, we see that dispersive edge states occur near $p_y=0$ with $\eta \le 1$. 

To reveal the nature of the dispersive edge states, we fix $\alpha_R$ and varies $\Delta_p$. As illustrated in  Fig.~\ref{Fig7} (a)-(c), when the amplitude $\Delta_p$ increases, $\pm$ nodal points from different quadrants move toward each other and annihilate. Correspondingly, as shown in (d)-(f), the region with dispersive edge modes around $p_y=0$ expands and the region with fermionic Andreev bound states shrinks. It is clear that dispersive edge modes near $p_y=0$ evolve into helical Majorana modes of the fully gapped $p$-wave superconductors. Therefore, when $\Delta_p > \alpha_R$, dispersive edge states around $p_y=0$ emerge and one expects that they are of the same nature as helical Majorana modes of the fully gapped $p$-wave superconductors.  

\begin{figure}[htp]
\includegraphics[width=8cm]{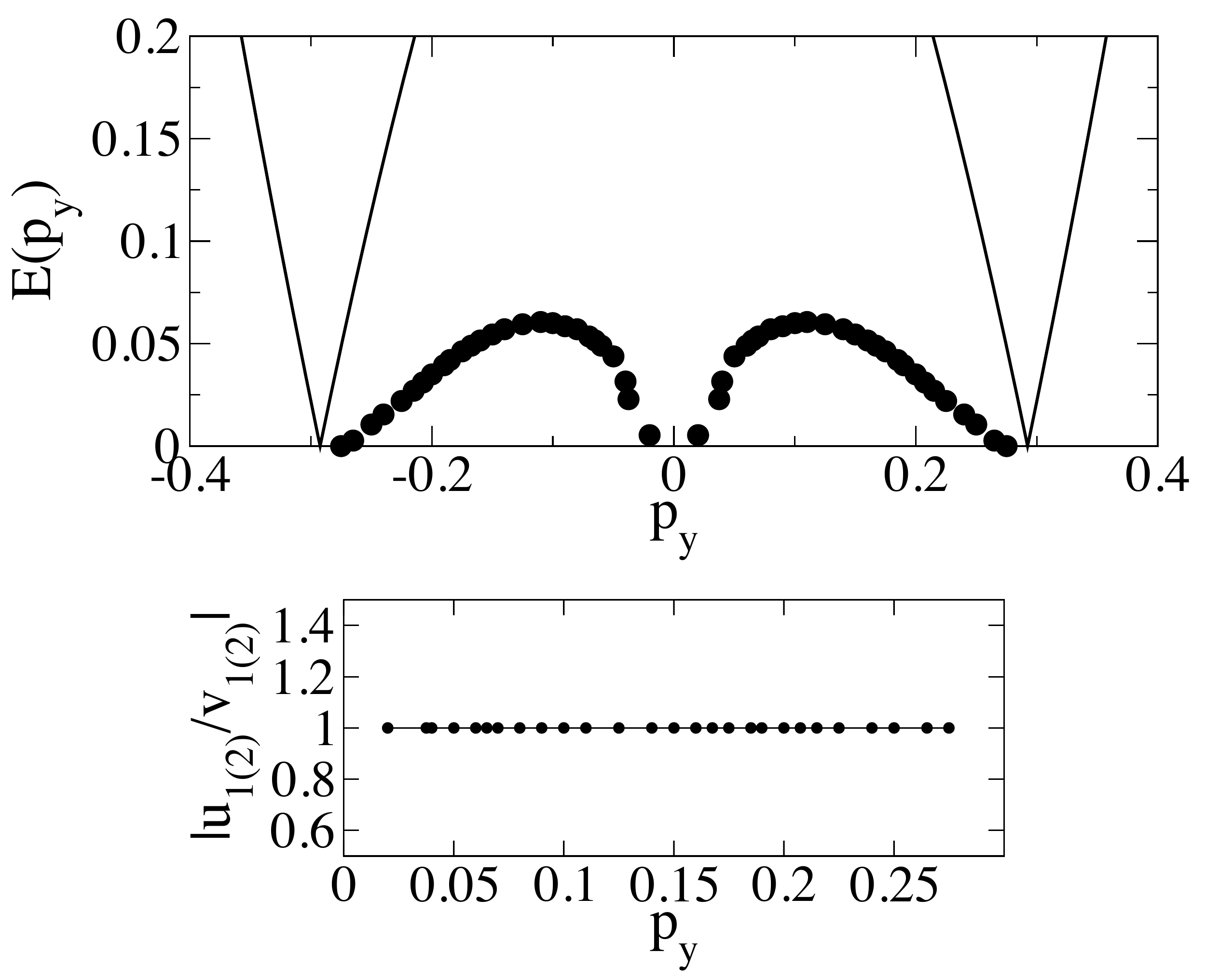}
\caption{Energy dispersions for $(110)$ edge near $p_y=0$. Here solid lines represent bulk states. Filled circles represent edge states. Lower panel: The Majorana condition, $\gamma^{\dagger}(E,p_y)=\gamma(-E,-p_y)$, is checked by examining $|v_1/u_1|=|v_2/u_2|=1$. Clearly, the dispersive edge states are Majorana modes.}
\label{Fig9}
\end{figure}

To examine if these dispersive edge mode are indeed Majorana modes, one needs to examine if the quasi-particle obeys $\gamma^{\dagger}(p_y,E)=\gamma(-p_y,-E)$\cite{Beenakker}. In general, the eigenstate at $i$ site
to the one-dimensional Hamiltonian $H_{1D}(p_y)$ is represented by a 4-component
$(v_1(p_y,E,i),v_2(p_y,E,i),u_1(p_y,E,i),u_2(p_y,E,i))$ so that the quasi-particle is given by 
\begin{eqnarray}
\gamma^{\dagger}(p_y,E,i)=u_1(p_y,E,i)c^{\dagger}_{p_y,i \uparrow}+u_2(p_y,E,i)c^{\dagger}_{p_y,i \downarrow} \nonumber \\
+v_1(p_y,E,i)c_{-p_y,i \uparrow}+v_2(p_y,E,i)c_{-p_y, i \downarrow}.
\end{eqnarray}
For Majorana modes, one requires $\gamma^{\dagger}(p_y,E,i)=\gamma(-p_y,-E,i)$, which in turn requires
$v_1(p_y,E,i)=u^{*}_{1}(-p_y,-E,i)$ and $v_2(p_y,E,i)=u^{*}_{2}(-p_y,-E,i)$. However, due to an uncertainty in defining $u_i$ and $v_i$ up to a $U(1)$ global phase factor, in order for $\gamma$ to be Majorana modes, one requires 
\begin{eqnarray}
v_1(p_y,E,i)=e^{i\phi} u^{*}_{1}(-p_y,-E,i), \nonumber \\
v_2(p_y,E,i)=e^{i\phi} u^{*}_{2}(-p_y,-E,i).
\end{eqnarray}
Therefore, $|v_1/u_1|=|v_2/u_2|=1$ is the condition for edge modes to be Majorana modes. In Fig.~\ref{Fig9}, we examine energy dispersions for $(110)$ edge near $p_y=0$. As shown in the inset, $|v_1/u_1|=|v_2/u_2|=1$ is satisfied for dispersive edge modes. Hence when $\Delta_p > \alpha_R$, the emerging dispersive edge states around $p_y=0$ are dispersive Majorana modes. 

\section{Tunneling Spectroscopy}
\label{tunnel}
The tunneling spectroscopy is one of means to examine Majorana fermions.  A typical experiment is
the measurement of the tunneling spectrum for normal metal $(N)$, insulator $(I)$ and superconductor $(S)$ junction, i.e., the NIS junction, in which zero-energy edge states would appear as a zero-bias conductance peak.  However, the zero-bias conductance peak may also arise from other mechanism\cite{Beenakker} and hence it is not considered as a smoking-gun evidence for Majorana fermions. On the other hand, the $4 \pi$ periodicity of supercurrent flowing across the Josephson junction is considered as a smoking-gun evidence for existence of Majorana fermions. In this section, we shall examine both the spectrum of the NIS junction and Josephson junction for high $T_c$ cuprates in the presence of Rashba spin-orbit interaction. 
It will be shown that dispersive Majorana modes would exhibit a unique feature as a small plateau in tunneling spectrum near zero bias peak. Furthermore, we find that the flat-band Majorana modes always result in 4$\pi$-Josephson effect in typical Josephson junctions. Therefore, by suitable designing  junctions involved in the tricrystal experiments, 
the flux trapped in the center of the tricrystal jumps in unit of two flux quanta, which would be a 
convincing way to detect Majorana fermions.

\subsection{Tunneling spectrum of NIS junction}
\label{tunnelspectrum}

The well-known zero-bias conductance peak observed in high-$T_c$ cuprates along $(110)$ direction manifests the fermionic Andreev bound states at the interface of NIS junctions\cite{Tanaka}. As indicated in the Sec.~\ref{gapless} C, in the presence of the Rashba spin-orbit interaction in high-$T_c$ cuprates, $(110)$ edge also hosts Majorana flat bands when $p$-wave is small. The detection of this Majorana flat bands is thus ambiguous in the NIS planar junctions due to the coexistence of zero-energy flat bands formed by fermionic and Majorana edge states. Both of them contribute to the zero-bias peak based in the Blonder-Tinkham-Klapwijk theory \cite{BTK}. On the other hand,  the dispersive Majorana fermions occur at finite energy scale and hence its appearance would be unambiguous. 

For a planar NIS junction, after partial Fourier transformation along the edge of the junction with $p_y$ being the wavevector along the edge, the junction is effectively a one-dimensional system.
For a give $p_y$, the effective one-dimensional Hamiltonian for the superconducting side with $(110)$ edge is given by Eq.(\ref{H1D}), $H_{1D} (p_y)$. Following Blonder, Tinkham, and Klapwijk\cite{BTK}, the NIS junction is formed by connecting $H_{1D} (p_y)$ with the effective 1D Hamiltonian of the normal metal side. If the one dimensional eigenstate $\hat{\Psi}(p_y)$ at $j$ site is represented by a 4-component $\hat{\psi}_{j}$. The tight-binding Bogoliubov equation can be 
generally written as
\begin{eqnarray}
&& \hat{T} \hat{\psi}_{j-1} + \hat{T}^{\dagger} \hat{\psi}_{j+1}+\hat{T}^{\prime} \hat{\psi}_{j-2}
+\hat{T}^{\prime\dagger} \hat{\psi}_{j+2}+\hat{\mu} \hat{\psi}_{j} \nonumber \\
&& + \hat{V}\delta_{x_j,-1} \hat{\Psi}_{j}=E \hat{\psi}_{j}, \label{Bogoliubov}
\end{eqnarray}
where the potential $\hat{V}=V\tau_z\otimes\sigma_0$ is the tunneling barrier at the interface due to the insulator layer, the superconductor $S$ occupies $j \geq 0$ and the normal metal $N$ occupies $j<0$. Matrices $\hat{T}$ and $\hat{T}^{\prime}$ denotes effective nearest neighbor and next nearest neighbor hopping, while $\hat{\mu}$ represents the effective on-site potential. They are given by
\begin{widetext}
\begin{eqnarray}
\nonumber&&\hspace{1cm}\hat{T}=\left (
\begin{array}{cccc}
 t_{\parallel} & t_{2} & \Delta_{t1} & \Delta_s\\
 t^{*}_{1} & t_{\parallel} & -\Delta_s & \Delta_{t2} \\
-\Delta_{t2} &-\Delta_s & -t_{\parallel}&  t^{*}_{1} \\
\Delta_s & -\Delta_{t1} & t_{2} & -t_{\parallel}
\end{array}\right ),\nonumber\hat{\mu}=\left (
\begin{array}{cccc}
 t_{\bot} & 0 & 0 & 0\\
0 & t_{\bot} & 0 & 0 \\
0 & 0 & -t_{\bot}& 0 \\
0 & 0 & 0 & -t_{\bot}
\end{array}\right ),
\hat{T}^{\prime}=\left (
\begin{array}{cccc}
 t^{\prime}_{\parallel} & 0 & 0 & 0\\
0 & t^{\prime}_{\parallel} & 0 & 0 \\
0 & 0 & -t^{\prime}_{\parallel}& 0 \\
0 & 0 & 0 & -t^{\prime}_{\parallel}
\end{array}\right ).
\end{eqnarray}
\end{widetext}
Here for the metal side, corresponding matrices are denoted by $\hat{T}_N$ and $\hat{\mu}_N$. 
$\Delta_s$, $\Delta_{t1}$, $\Delta_{t2}$, and $t^{\prime}_{\parallel}$ are set to zero, while
$ t_{\parallel}$, $ t_{\bot}$, $t_1$, and $t_2$ are set to be hopping amplitudes $-2t_N \cos(p_y\delta_{1y})$, $-\mu_N$,
$\frac{\alpha_{R}}{2}(-ie^{ip_y\delta_{1y}}+e^{-ip_y\delta_{1y}})$, and  $\frac{\alpha_{R}}{2}(ie^{-ip_y\delta_{1y}}-e^{ip_y\delta_{1y}})$.
For the superconducting side, corresponding matrices are denoted by $\hat{T}_S$, $\hat{\mu}_S$ and $\hat{T}_S^{\prime}$ with $ t_{\parallel}=-2t\cos(p_y\delta_{1y})$, $t^{\prime}_{\parallel}=-t^{\prime}-2t^{\prime\prime}\cos(p_y\delta^{\prime\prime}_{1y})$, $t_{\bot}=-2t^{\prime}\cos(p_y\delta^{\prime}_{2y})-\mu$, $t_{1}=\frac{\alpha_{R}}{2}(-ie^{ip_y\delta_{1y}}+e^{-ip_y\delta_{1y}})$, $t_{2}=\frac{\alpha_{R}}{2}(ie^{-ip_y\delta_{1y}}-e^{ip_y\delta_{1y}})$, $\Delta_s=-i\Delta_d\sin(p_y\delta_{1y})$, $\Delta_{t1}=\frac{\Delta_p}{2}(ie^{-ip_y\delta_{1y}}+e^{ip_y\delta_{1y}})$ and $\Delta_{t2}=\frac{\Delta_p}{2}(-ie^{-ip_y\delta_{1y}}+e^{ip_y\delta_{1y}})$.
\begin{figure}[h]
\hspace{0.5cm}\includegraphics[width=9cm]{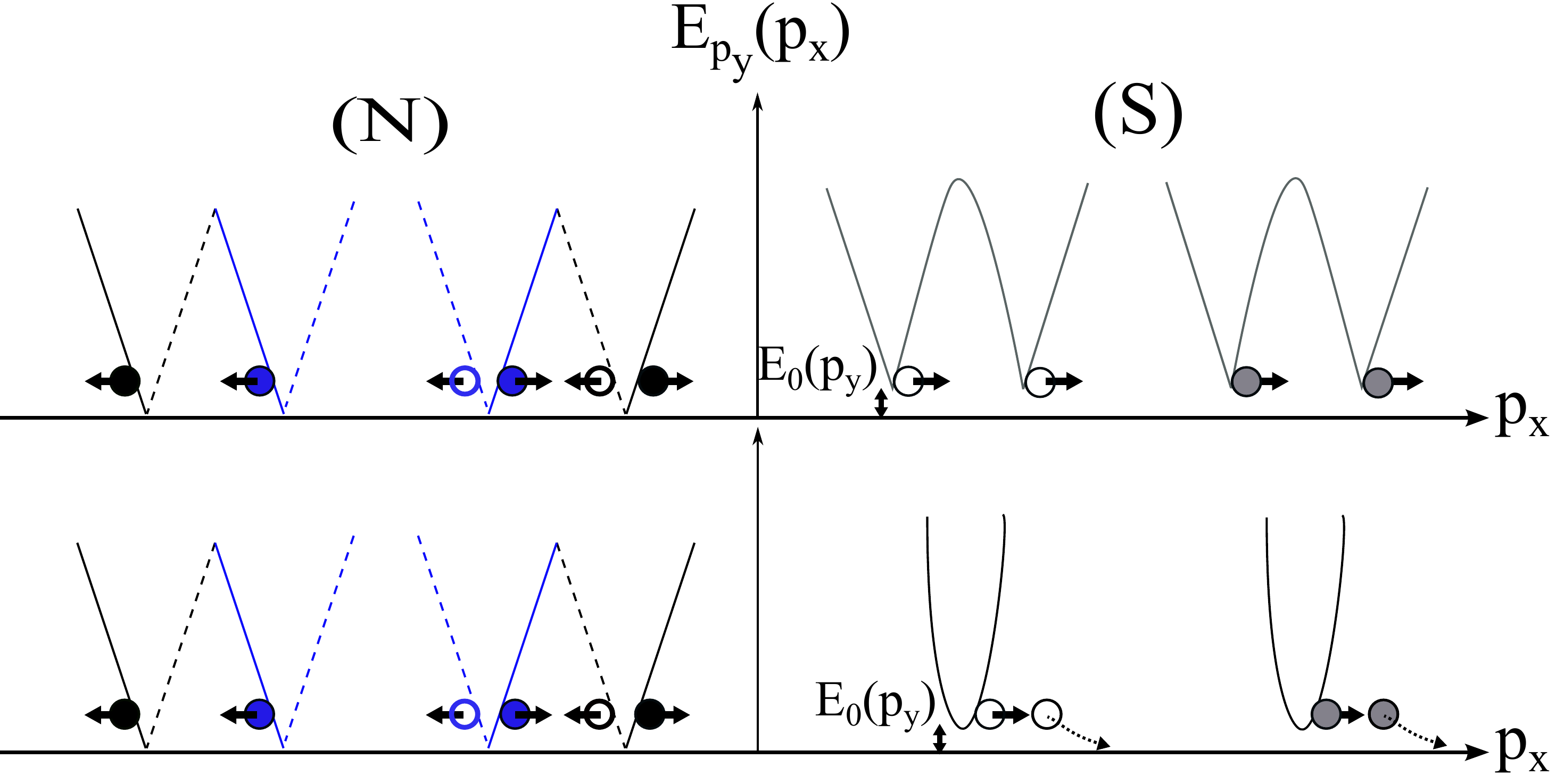}\caption{
Schematic diagrams of the NIS junction for quasi-particles traveling (right arrows) from N to S and 
reflected (left arrows) from S to N.
Here upper panel is for $\alpha_R > \Delta_p$ and lower panel is for $\alpha_R < \Delta_p$.
In the N side, solid lines and dash lines represent spectra of particles and holes respectively.
Different colors (black and blue) stand for two bands split by the Rashba interaction. 
Black and blue dots with arrows represent incident and reflected particles, while open circles denote the reflected holes.
In the S side for $E>E_0(p_y)$, gray dots stand for the particle-like and open circles denote the hole-like transmitted modes.
For $\alpha_R<\Delta_p$, only two nodal points are present in $S$ side. Specifically, two evanescent modes represented by dash arrows can be found  by solving $E(p^n_x,p_y)=E$ with $E>E_0(p_y)$. Here, $E(p^n_x,p_y) $ is the lowest positive $E_p$ in Eq.(\ref{Ek}).}
\label{Fig10}
\end{figure}   

At a given energy $E$, far away from the interface, the quasi-particle wavefunction $\hat{\psi}_{j}$ satisfies the bulk Bogoliubov equation. As illustrated in Fig.~\ref{Fig10}, there are two incident waves (indicated as black and blue dots) in the N side, which will be denoted by $\Phi_{\bar{e}_1}$ and $\Phi_{\bar{e}_2}$. 
For reflected waves, there are 2 bulk quasi-particle wavefunctions indicated by black and blue dots and 2 bulk quasi-hole wavefunctions indicated by open circles. The corresponding normalized wavefunctions are denoted by $\Phi_{e_1}$, $\Phi_{e_2}$, $\Phi_{h_1}$,  and $\Phi_{h_2}$. Similarly, for the S side, there are 2 particle-like and 2 hole-like wavefunctions denoted as  $\phi_{e_1}$, $\phi_{e_2}$, $\phi_{h_1}$,  and $\phi_{h_2}$. By using these wavefunctions, 
$\hat{\psi}_j(p_y)$ can be expressed as
\begin{eqnarray}
\text{for $j \geq 0$},  &&\hat{\psi}_j(p_y) = \sum_{n=e_1,e_2,h_1,h_2} t_n z^j_{n} \phi_n (p_y,z_n), \nonumber \\
\text{ for $j<0$},  && \hat{\psi}_j(p_y) \equiv \hat{\Psi}_j( p_y) \nonumber \\
           &&= \sum_{n=\bar{e}_1,\bar{e}_2, e_1,e_2,h_1,h_2} r_n \bar{z}^j_n \Phi_n (p_y,z_n). 
 \label{psi}
\end{eqnarray}
Here indices with $\bar{e}_1$ and $\bar{e}_2$ represent quantities of incident waves. The  wavevectors $p^n_x$ and $\bar{p}^n_x$ in the corresponding quantities $z_n\equiv e^{ip^n_x}$ and $\bar{z}_n\equiv e^{i \bar{p}^n_x}$ are perpendicular to the interface in the $S$ side and the $N$ side respectively. $p^n_x$ and $\bar{p}^n_x$ must be solved from
the bulk energy dispersion $E(p^n_x,p_y)=E$. 
$t_n$ and $r_n$ are corresponding transmission and reflection amplitudes for $\phi_n (p_y,z_n)$ and $\Phi_n (p_y,z_n)$.  Solutions for $j \geq 0$ and $j<0$ with forms given by Eq.(\ref{psi}) are then joined together at the interface $j=0$. By comparing Eq.(\ref{Bogoliubov}) at $j \gg 0$ and $j \ll 0$ with Eq.(\ref{Bogoliubov}) at $j = -1$ and $j =0 $, we find that the boundary conditions are given by
\begin{eqnarray}
&&\hat{\psi}_0=\hat{\Psi}_0-(\hat{T}^{\dagger}_N)^{-1} \hat{V}\hat{\Psi}_{-1}, \nonumber \\ 
&&\hat{T}^{\dagger}_N \hat{\Psi}_{-1}=\hat{T}^{\dagger}_S \hat{\psi}_{-1}.\label{IV-1}
\end{eqnarray}
After re-arrangements of the above two boundary conditions, amplitudes $t_n$ and $r_n$ can be solved by the following $8\times8$ linear equation
\begin{widetext}
\begin{eqnarray}
\nonumber &\left(\begin{array}{cccccccc} \phi_{e_1} & \phi_{e_2} & \phi_{h_1} & \phi_{h_2} &\hat{\mathcal{U}}_{e_1} \Phi_{e_1}&\hat{\mathcal{U}}_{e_2} \Phi_{e_2}&\hat{\mathcal{U}}_{h_1} \Phi_{h_1}&\hat{\mathcal{U}}_{h_2} \Phi_{h_2}\\
\hat{T}^{\dagger}_S z^{-1}_{e_1} \phi_{e_1}&\hat{T}^{\dagger}_S z^{-1}_{e_2} \phi_{e_2}&\hat{T}^{\dagger}_S z^{-1}_{h_1} \phi_{h_1}&\hat{T}^{\dagger}_S z^{-1}_{h_2} \phi_{h_2}&-\hat{T}^{\dagger}_N \bar{z}^{-1}_{e_1} \Phi_{e_1}&-\hat{T}^{\dagger}_N \bar{z}^{-1}_{e_2} \Phi_{e_2}&-\hat{T}^{\dagger}_N\bar{z}^{-1}_{h_1} \Phi_{h_1}&-\hat{T}^{\dagger}_N\bar{z}^{-1}_{h_2} \Phi_{h_2} \end{array}\right)&\\&\cdot
\left(\begin{array}{cccccccc}t_{e_1} &t_{e2} &t_{h_1} &t_{h_2} &r_{e_1}&r_{e_2}&r_{h_1}&r_{h_2}\end{array}\right )^{T}
=\left(\begin{array}{cc}-\hat{\mathcal{U}}_{\bar{e}_1} \Phi_{\bar{e}_1}-\hat{\mathcal{U}}_{\bar{e}_2} \Phi_{\bar{e}_2}\\
\hat{T}^{\dagger}_N(\bar{z}^{-1}_{\bar{e}_1} \Phi_{\bar{e}_1}+\bar{z}^{-1}_{\bar{e}_2} \Phi_{\bar{e}_2})\end{array}\right), &
\label{Andreev_eq}
\end{eqnarray}   
\end{widetext}
where $\hat{\mathcal{U}}_n =( \hat{T}^{\dagger}_N)^{-1} \hat{V} \bar{z}^{-1}_n -1$.

For a given $p_y$, there is a minimum of bulk energy at $E_0(p_y)$ as illustrated in Fig.~\ref{Fig10}. 
When the energy $E$ of an injected particle with $E>E_0(p_y)$, $p^n_x$ and $\bar{p}^n_x$ are both real. We obtain two transmitted electron-like and two hole-like quasiparticles in the $S$ side and two reflected electrons and holes in the $N$ side. On the other hand, for $E<E_0(p_y)$, two electron-like and two hole-like in the $S$ side become evanescent modes with $p^e_x$ and $p^h_x$ being complex numbers. Note that for $\alpha_R <\Delta_p$, even for $E>E_0(p_y)$, since there are only two nodal points, both transmitted and evanescent modes are present in the $S$ side. 

Solving Eq. (\ref{Andreev_eq}), one obtains solutions for $t_n$ and $r_n$ which then determine the transmitted and reflected particle currents given by 
\begin{eqnarray}
\vec{J}_{j,j^{\prime}}&=&i\left(\hat{\Psi}^{\dagger}_{j^{\prime}}\hat{T}^{\dagger}_N\hat{\Psi}_j-
\hat{\Psi}^{\dagger}_{j}\hat{T}_N\hat{\Psi}_{j^{\prime}}\right),
\end{eqnarray}
in the $N$ side and similar expression for the $S$ side with $\hat{T}_N$ and $\hat{\Psi}_j$ replaced by $\hat{T}_S$ and $\hat{\psi}_j$. Hence we obtain the particle current $J^e_t$ through the tunneling barrier $\hat{V}$ in the $N$ side 
and the reflected particle and hole currents: $J^e_{r}$ and $J^h_{r}$.  The normalized tunneling conductance for the NIS junction is then given by\cite{BTK}
\begin{eqnarray}
\sigma_N(E,V) =\frac{1}{g_0(E,V)}\sum_{p_y}\sigma_{n}(E,V,p_y), 
\end{eqnarray}
where $g_0(E,V) = \sum_{p_y} J^e_{t}(E,V,p_y) / J^e_{i}(E,p_y)$ is the ballistic conductance of the $N$ side and $\sigma_n(E,V,p_y)$ is defined to be $1-[ J^e_{r}(E,V,p_y)-J^h_{r}(E,V,p_y) ]/J^e_{i}(E,p_y)$.

\begin{figure}[htp]
\begin{center}
\includegraphics[width=8.0cm]{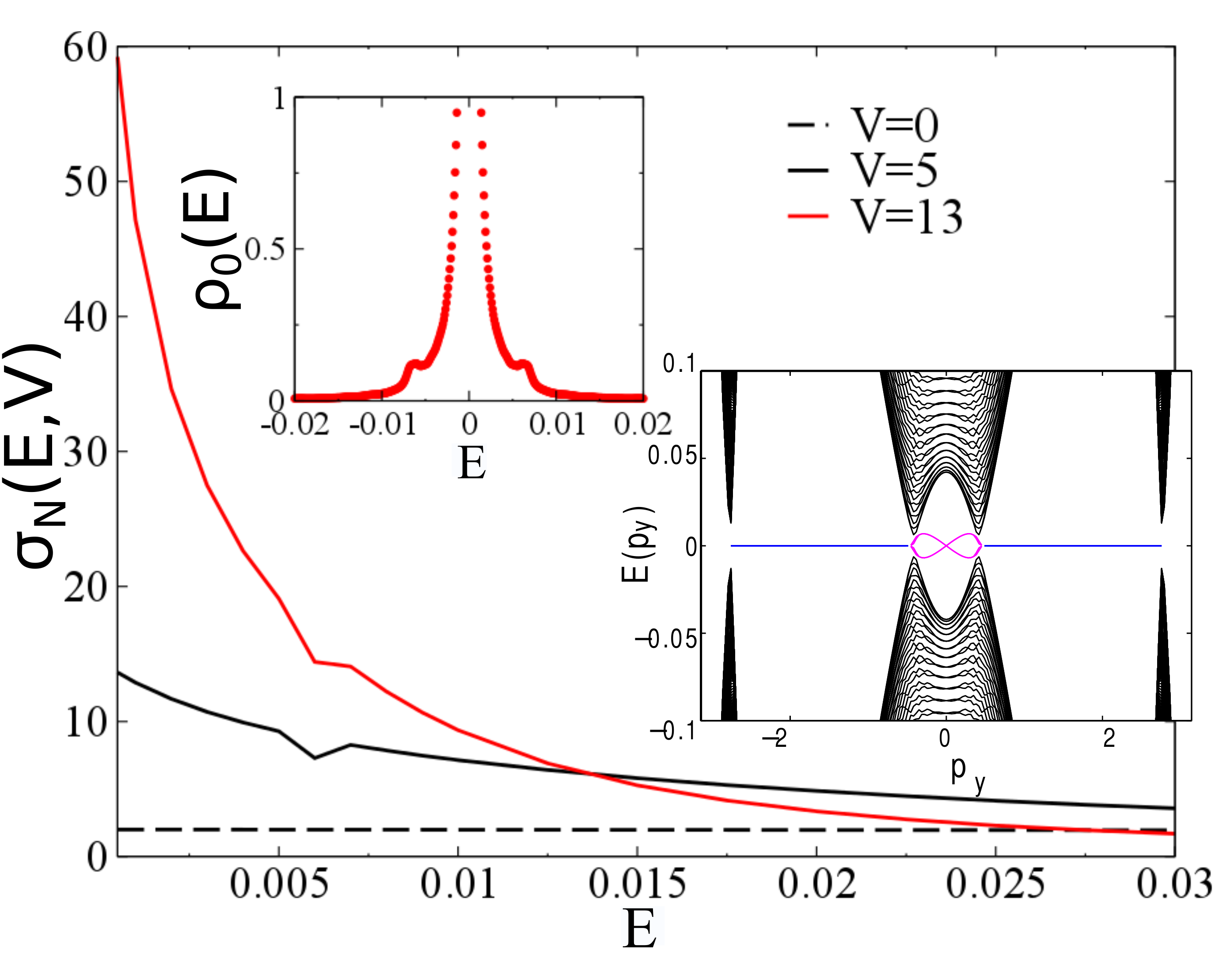}
\caption{
Tunneling conductance near zero bias for Metal-Insulator-high $T_c$ cuprate superconductor in the presence of Rashba spin-orbit interaction ($\alpha_R < \Delta_p$). Here $E$ is measured in terms of $t$ with $t^{\prime}=t^{\prime\prime}=0,\alpha_R=0.13,\Delta_d=0.32,\Delta_p=0.16$.
The tunneling conductance $\sigma_N$ near $E=0$ exhibits a candleholder-like structure, which reflects the local density of states at the edge in the strong interface scattering limit ($V \rightarrow \infty$).
The insets show the corresponding local density of states and the energy spectrum in which 
the Andreev flat edge modes contributes to the zero-energy divergence while the dispersive edge modes contribute to the plateau.}
\label{Fig11}
\end{center}
\end{figure}

In Fig.~\ref{Fig11}, we show the computed tunneling conductance. In the strong interface scattering limit $V\rightarrow \infty$, the tunneling conductance is proportional to local density of state $\rho_0(E)$ at $(110)$ edge\cite{BTK}, dominated by edge states. 
For edge states with the dispersion $E (p_y) \propto p^{m}_y$, the local density state is $\rho_0(E) \propto E^{\frac{1-m}{m}}$ as illustrated in the inset of Fig.~\ref{Fig11}.
Clearly, the zero-bias conductance peak in Fig.~\ref{Fig11} manifests the dispersionless Andreev bound states shown in the inset of Fig.~\ref{Fig11} while the small plateau at finite energy scale reveals the linearly dispersive Majorana fermion near $p_y=0$. As a consequence, an overall candleholder-like structure of the tunneling conductance reflects the coexistence of Andreev flat-band and dispersive Majorana fermions at the interface.  
  
\subsection{SIS$'$-Josephson Junction}

In this subsection, based on our mean-field solutions, the quasi-particle spectrum and Josephson current in SIS$'$ junctions are investigated. We shall examine periodicity of Josephson current in SIS$'$ junctions with S and S$'$ formed by high $T_c$ cuprates in different edge orientations as illustrated in Fig.~\ref{Fig13}(a). In particular, we will show that the hybridization of Majorana zero-energy flat-bands of S and S$'$ leads to $4\pi$-periodicity\cite{Kitaev}. However, we show that this only occurs when there is an overlap between Majorana zero-energy flat-bands from $S$ and $S'$. Based on these findings, we explore and design the tricrystal junction which can generate a spontaneous vortex with half-flux quanta with jump in unit of two flux quanta as the magnetic field is applied. This could serve as a way for detecting the elusive Majorana fermions in the high-$T_c$ cuprate-based heterostructure.  

We start by representing the wavevector along the junction by $p_y$. For a given $p_y$, the thin insulating layer  $I$ and superconducting sides S and S$'$ are effectively one-dimensional with coordinates denoted by $x$ and their corresponding Hamiltonians are generally given by Eq.(\ref{H1D}), denoted by $H_I (p_y)$, $H_S (p_y)$, and $H_{S'}(p_y)$ respectively.  The effective Hamiltonian of SIS$'$  is then given by
\begin{eqnarray}
H_{SIS'} (p_y) &=& H_S (p_y) +H_I (p_y) + H_{S'} (p_y) \nonumber  \\
            &+& H_{SI} (p_y) +  H_{IS'}(p_y) + H.C., \label{SIS}
\end{eqnarray}
where the insulator occupies $-x_N<x<x_N$, S occupies $x \leq -x_N$, S$'$ occupies $x \geq x_N$, and $H_{SI}(p_y)$ and $H_{IS'}(p_y)$ are the tunneling Hamiltonians  describing tunneling between superconductors and the thin insulator.  To investigate the Josephson current, we impose a a $U(1)$ phase difference $\phi=\phi_{S'}-\phi_S\equiv\phi$ between the pairing potentials in S and S$'$ sides by setting the pairing potential $\hat{\Delta}_S=\hat{\Delta} (p_y)\theta(-x_i-x_N)$ for S side and  $\hat{\Delta}_{S'}=\hat{\Delta} (p_y)e^{i\phi}\theta(x_i-x_N)$ for S$'$ side. 
\begin{figure}[htp]
 \begin{center}
\includegraphics[width=8.5cm]{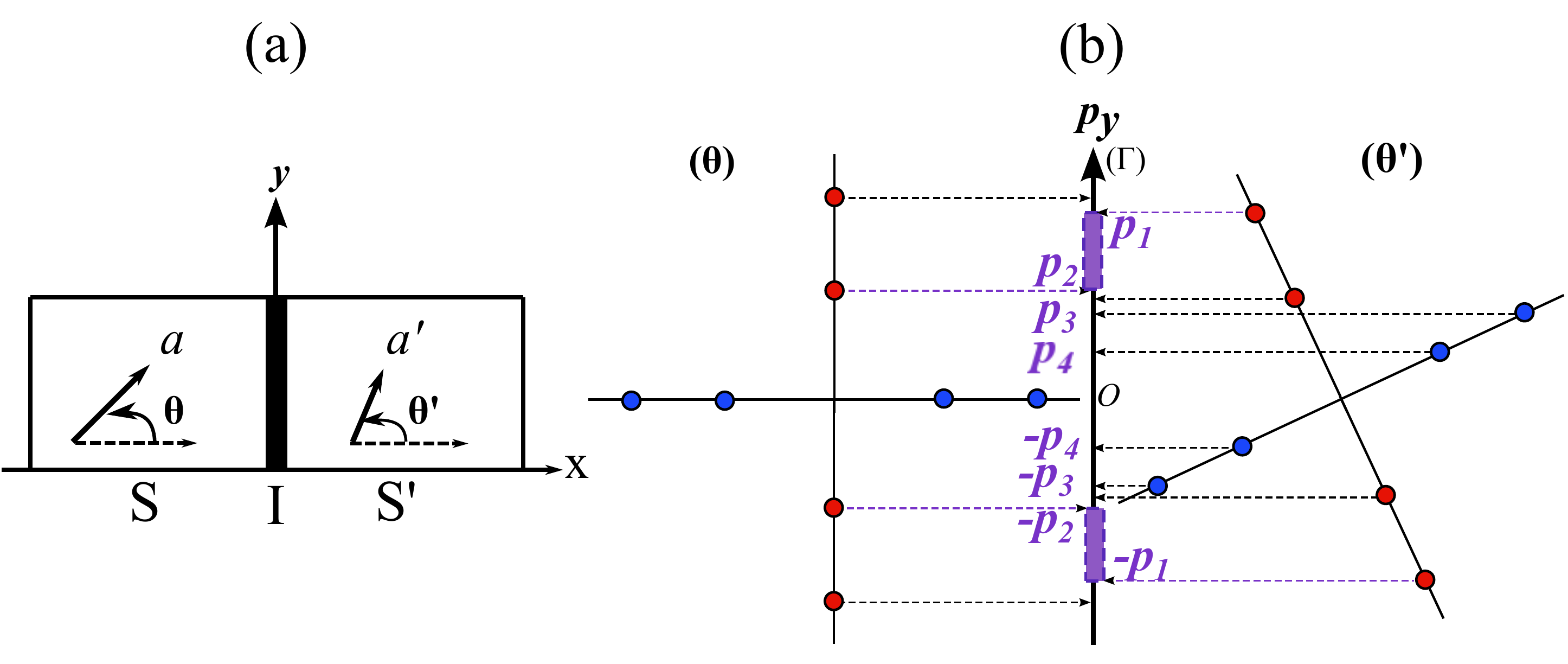}\caption{(a) Schematic plot of a SIS$'$ junction. Here the relative angles of $(100)$ directions (denoted by $a$ and $a'$ axes) of S and S$'$ to the tunneling direction ($x$ axis) are $\theta$ and $\theta'$ respectively. (b) Comparison of projected $p_y$ along the junction for edge states that would appear in isolated and semi-infinite S and S$'$. Here $\theta' \geq \theta$  and red and blue dots are nodal points. It is seen that $p_y$'s of Majorana zero-energy modes (flat-bands) from S and S$'$ may overlap in the
region $p_2 \leq |p_y| \leq p_1$ and result in $4 \pi$ periodicity in Josephson current.}\label{Fig12}
\end{center}
\end{figure}   

We first analyze edge states of isolated semi-infinite superconducting sides (S and S$'$). Following Fig.~\ref{Fig6}, by projecting nodal structures onto an edge, one obtains possible number of edge modes for each $p_y$ along the edge. In Fig.~\ref{Fig12}(b), we show the comparison of two edges with difference orientations. It is seen that a junction formed by these two edges may hybridize two Majorana zero-energy flat-bands (e.g. $p_2 \leq |p_y| \leq p_1$), one fermionic Andreev flat-band with one Majorana zero-energy flat-band (e.g. $p_4 \leq |p_y| \leq p_3$), or two fermionic Andreev flat-bands (e.g. $|p_y| \leq p_4$). As we shall see, only the overlap of zero-energy Majorana bound states lead to their hybridization and result in $4 \pi$ periodicity of Josephson current when these two edges form a SIS$'$ junction\cite{Kitaev}. If the relative angle of $(100)$ directions of a superconductor S to the tunneling direction ($x$ axis) is $\theta$, the relative angle of projected $p_y$ to the $(100)$ direction is $\pi/2-\theta$. Hence projection of the Majorana zero-energy flat-band onto the junction interface is $k^- \cos(\pi/4 - \theta) \leq  p_y \leq k^+ \cos(\pi/4 - \theta)$, where $k^{\pm}$ are magnitudes of wavevectors for nodal points. 
Assuming that both S and S$'$ have the same doping level, the projection of Majorana zero-energy flat-band for S$'$ is also determined by $k^{\pm}$ and hence the projected Majorana zero-energy flat-band is given by $k^- \cos(\pi/4 - \theta') \leq  p_y \leq k^+ \cos(\pi/4 - \theta')$.  For doping $\delta \sim 0.16$ and $\alpha_R \sim 0.05-0.3$, we find that $k^-/k^+ \sim 0.92-0.77$ and the Majorana zero-energy flat-bands for the SIS$'$ junction overlap as long as 
\begin{eqnarray}
|\theta - \theta'|  \lesssim 21^\circ-39^\circ .\label{theta}
\end{eqnarray}

To illustrate effects of overlapped zero-energy Majorana flat-bands in the junction, we examine quasi-particle spectrum and Josephson current for junctions : (i) $\theta = 45^{\circ}$/$(1\bar{1}0)$ edge, $\theta'=0^{\circ}$/$(100)$ edge and (ii)  $\theta = 45^{\circ}$, $\theta'=\tan^{-1}2$/$(2\bar{1}0)$ edge. In Fig.~\ref{Fig13}(a), we show quasi-particle spectrum for the  S($45^{\circ}$)-I-S$'$($0^{\circ}$) junction. Here due to finite lengths of S ans S$'$ ($Nx=625$), in-gap states include edge states from outer edges of S and S$'$.  It is seen that there is no Majorana zero-energy state (single branch) formed by hybridization of edge states from S ans S$'$.  This is verified by checking energies of edge states versus
the $U(1)$ phase difference $\phi$ shown in  Fig.~\ref{Fig13}(b). The period is $2 \pi$. 
\begin{figure}[htp]
\includegraphics[width=4.5cm]{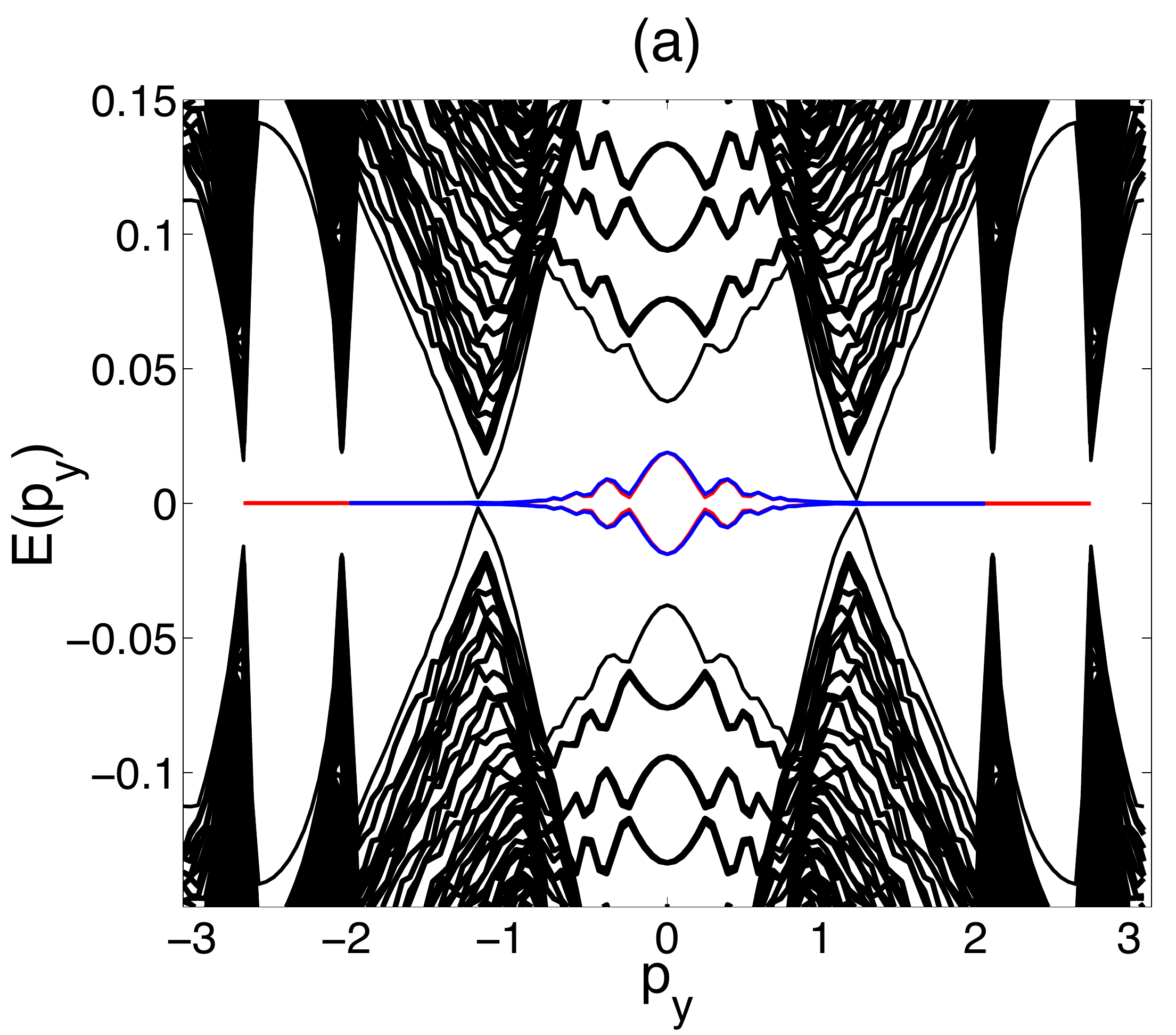}\includegraphics[width=4.5cm]{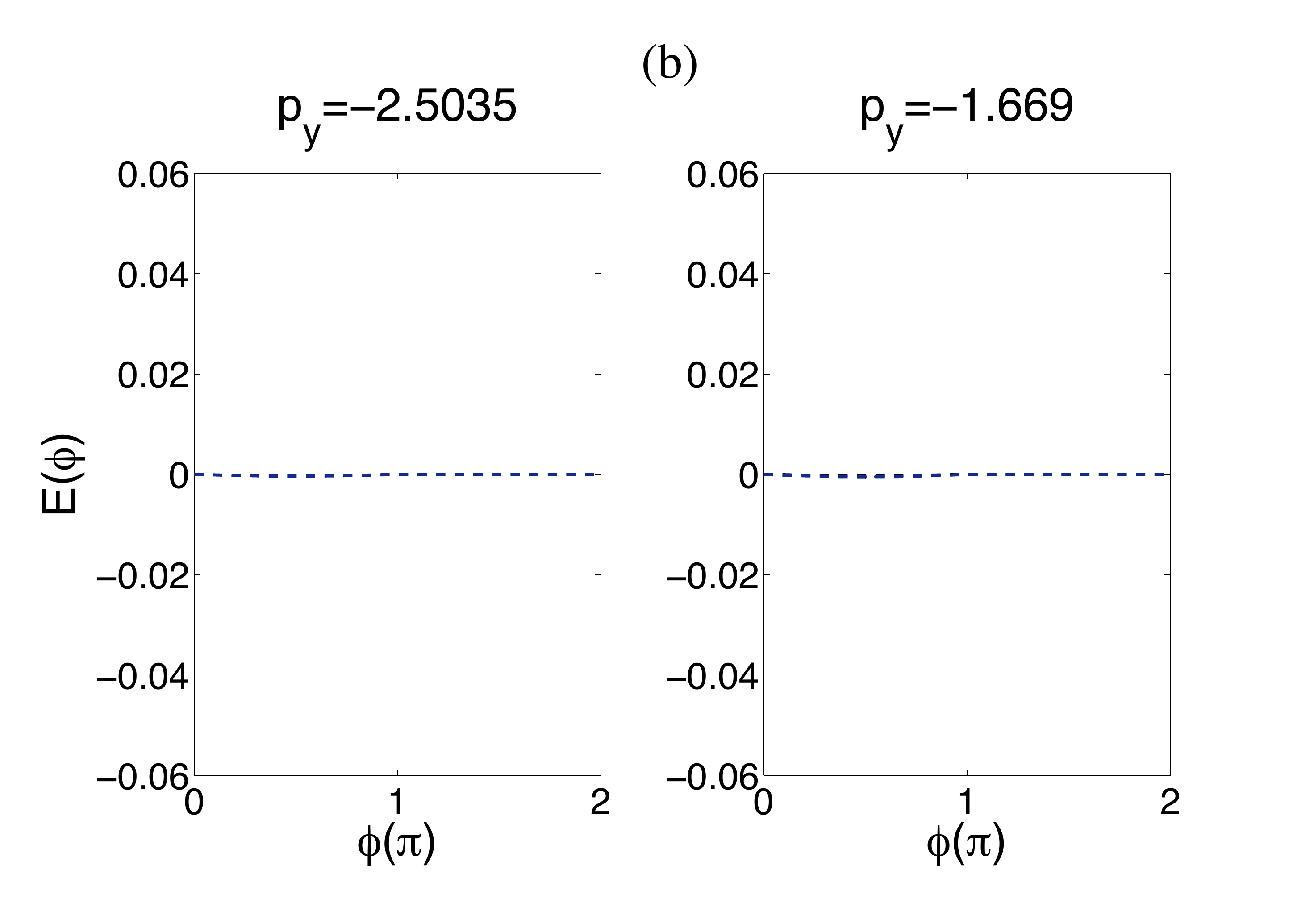}\caption{ (a) Low-lying quasi-particle spectrum for S($45^{\circ}$)-I-S$'$($0^{\circ}$) junction with $\phi=0.5\pi$. Here due to finite lengths of S ans S$'$($Nx=500$), in-gap states include edge states from outer edges of S and S$'$.  Edge states inside the junction are marked by blue (Andreev Fermionic modes) and red (Majorana modes) lines. It is seen that single branch Majorana flat-band is missing. (b) Energies versus the phase difference $\phi$ for edge states in the junction for a few $p_y$'s. The periods are found to be $2\pi$. Here length of S and $S'$ is $N_x=625$.}\label{Fig13}
\end{figure} 
Next, we consider the S($45^{\circ}$)-I-S$'$($\tan^{-1} 2$) junction. Fig.~\ref{Fig14}(a) shows 
the full quasi-particle spectrum. It is seen that there is a Majorana flat-band for $2 \lesssim |p_y| \lesssim 3$, resulting from hybridizing two Majorana zero-energy flat-bands from S and S$'$. In addition, there are in-gap states due to 
edge states in the junction formed by hybridizing one fermionic Andreev flat-band with one Majorana zero-energy flat-band  or two fermionic Andreev flat-bands. In Fig.~\ref{Fig14}(b), we examine energy versus phase difference $\phi$ for edge states in the junction.  It is clearly seen that only Majorana modes due to hybridization of Majorana fermions from S and $S'$ exhibit
$4\pi$ periodicity.  The $4 \pi$ period results from conservation of the fermion parity\cite{Beenakker,Liang} and the fermion parity is conserved only when two Majorana zero-energy modes are hybridized. We thus verified that only when the Majorana zero-energy flat-bands for each edge that form the junction overlap, the supercurrent in the junction shows $4 \pi$ periodicity.
\begin{figure}[htp]
\includegraphics[width=4.1cm]{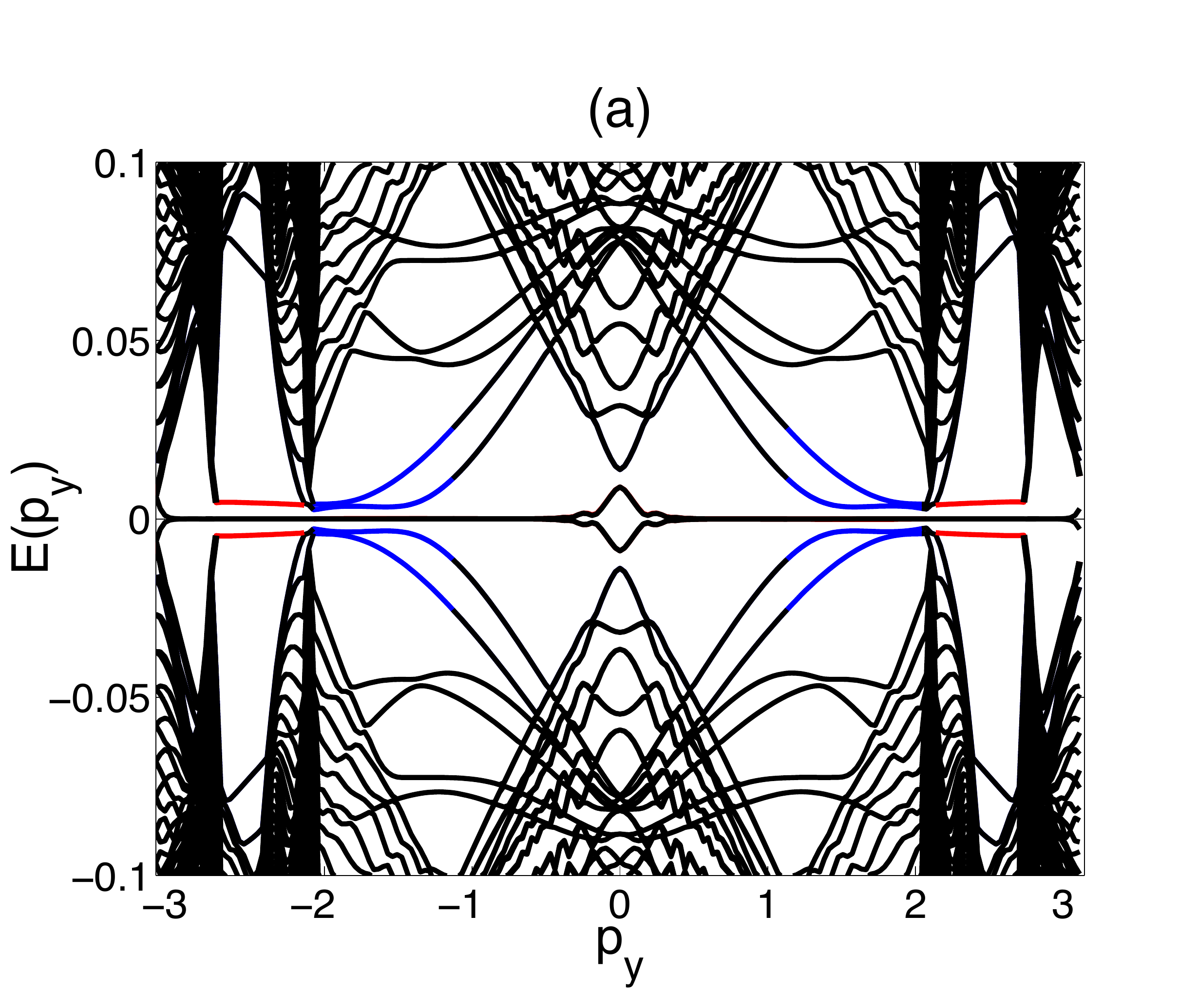}
\includegraphics[width=4.4cm]{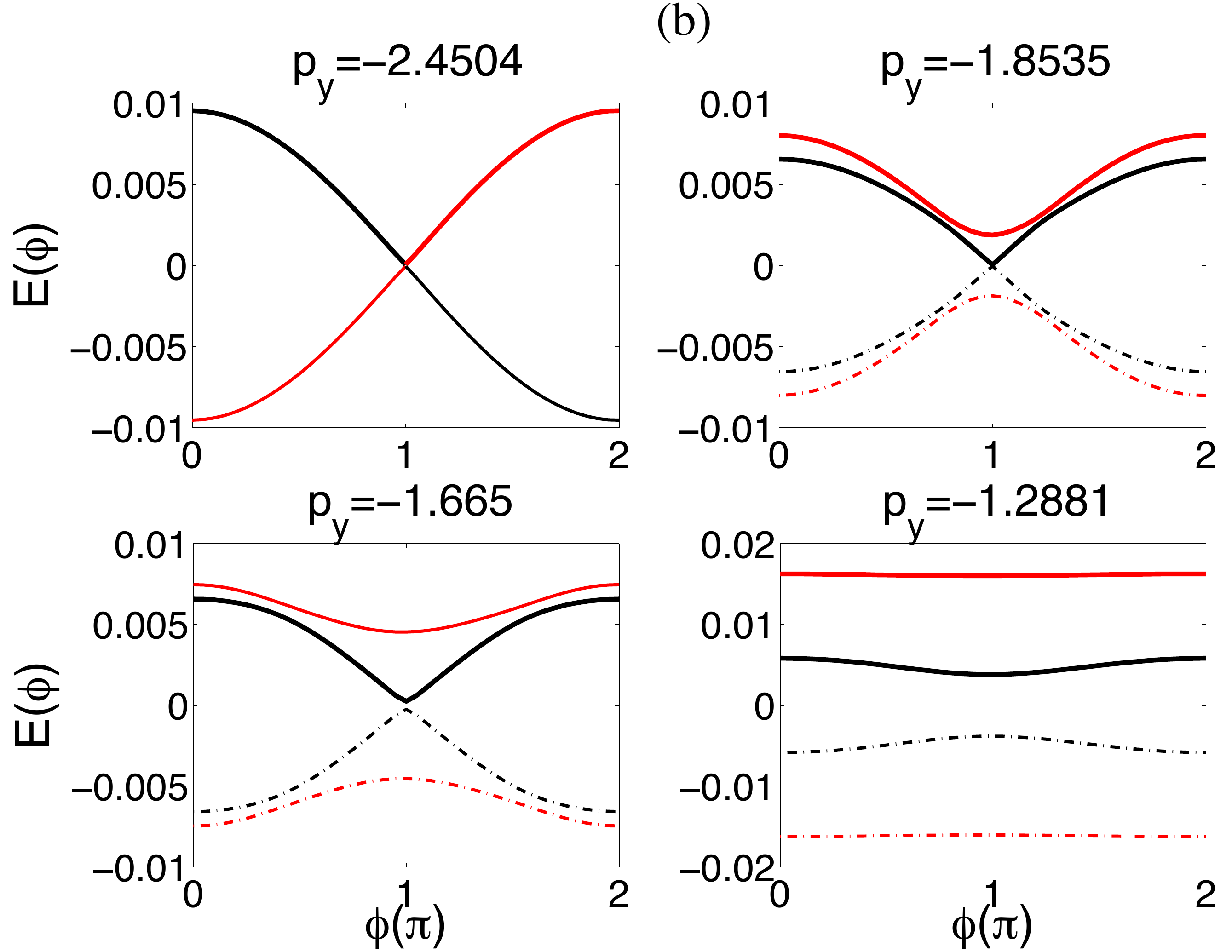}\caption{(a) Quasi-particle spectrum of S($45^{\circ}$)-I-S$'$($\tan^{-1} 2$) junction for $\phi =0.5 \pi$. Here length of S and S$'$ is $N_x=500$. Edge states inside the junction are marked by blue (Andreev fermionic modes) and red (Majorana modes) lines. (b) Energy versus phase difference $\phi$ for edge states in the junction for a few $p_y$'s. Here length of S and $S'$ is $N_x=625$. Plot of $p_y=-2.4504$ represents the spectrum due to the hybridization of Majorana fermions from S and $S'$. The red and black solid lines denote different fermion parity $P=\pm1$. Plots of $p_y=-1.8535$ and $p_y=-1.665$ stand for the spectrum formed by hybridizing fermionic Andreev zero-energy bound states. The remaining plot for $p_y=-1.2881$ is the spectrum resulted from hybridization of fermionic and Majorana zero energy modes. It is seen that only  hybridization of Majorana fermions results in $4\pi$ periodicity.
}\label{Fig14}
\end{figure} 

\begin{figure}[htp]
\includegraphics[width=6cm]{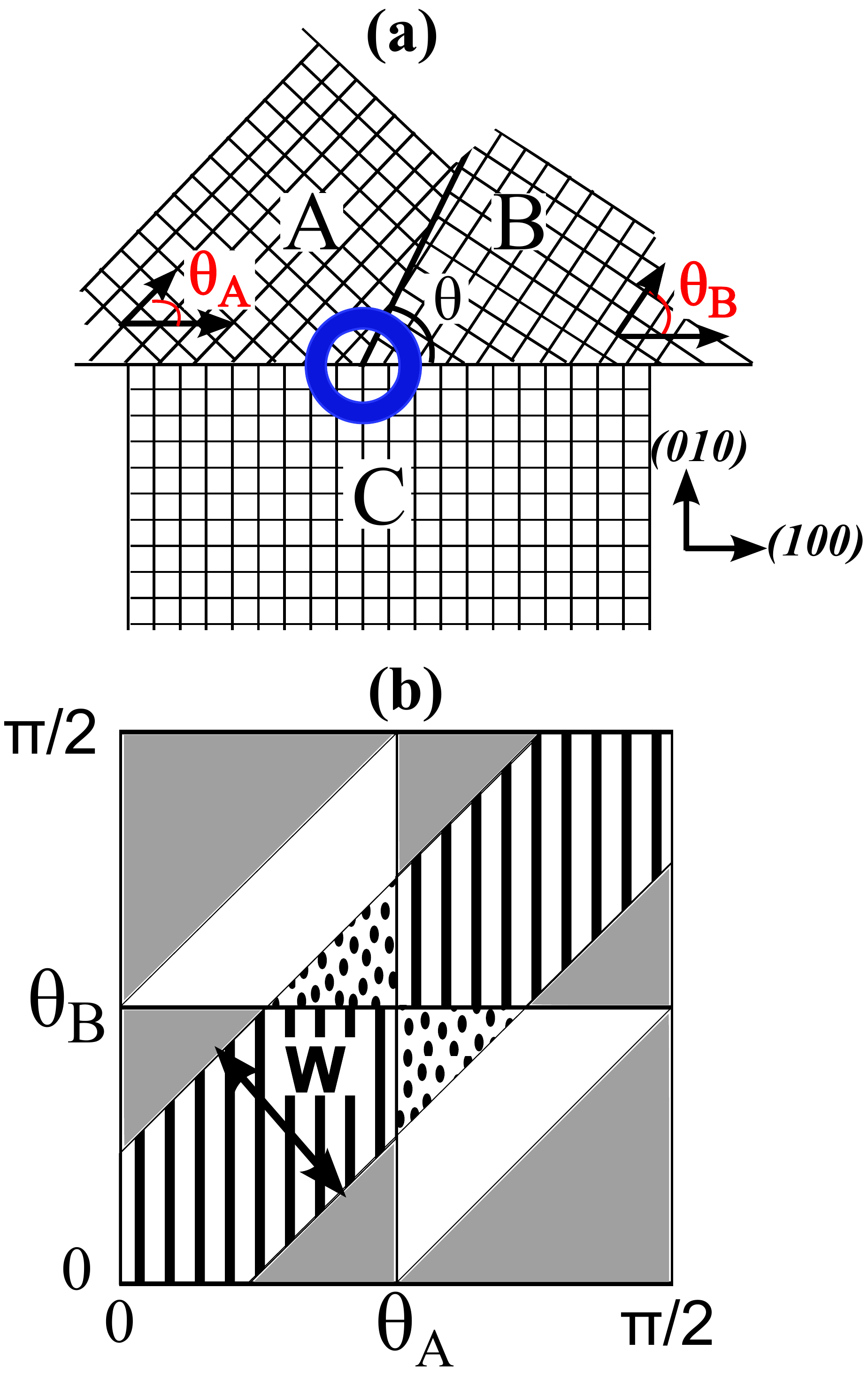}\caption{(a) Configuration of tricrystal experiment. Here $\theta_A$ and $\theta_B$ are the angles between the $a$ axis of the upper left crystal (A) and right crystal (B) and 
$(100)$ direction of the bottom crystal. (b) Regions of parameters for $\pi$ rings with Majorana modes (shaded areas with dots). Here $\theta \simeq \theta_B$. Grey areas are regions for zero ring\cite{Tsuei2}. White areas are regions for $\pi$ rings without Majorana modes, and shaded areas with lines are regions for zero rings with Majorana modes. For doping $\delta = 0.16$ and $\alpha_R = 0.05-0.3$, the width $W$ for the region with Majorana fermions is in the range of $42^\circ-78^\circ$.}\label{Fig15}
\end{figure} 

The $4 \pi$ periodicity implies fluxes of vortices trapped in the ring formed by the SIS$'$ junction jump in
unit of two flux quanta\cite{Beenakker}. One of configurations that allows one to explore the pairing symmetry
and also vortices trapped in high $T_c$ cuprates is the tricrystal experiment with a $\pi$-ring\cite{Tsuei1}. 
Here we examine the periodicity of $U(1)$ phase winding phase for the tricrystal configuration shown in Fig.~\ref{Fig15}(a).
As illustrated in Fig.~\ref{Fig15}(a), for general tricrystal configurations, one may fix the edge direction of the cuprate in the bottom labelled by C to be in $(010)$ direction. Let angles between $a$ axes of the upper left crystal (A) and 
right crystal (B) and $(100)$ direction of the bottom crystal be $\theta_A$ and $\theta_B$. If the interface between A and B be $\Gamma$ and the angle between $\Gamma$ and the $(100)$ direction be $\theta$, the interface $\Gamma$ is an S($\theta-\theta_A$)-I-S$'$($\theta-\theta_B$) junction. Clearly, since there is no Majorana zero-energy mode in the $(010)$ edge, in the tricrystal configuration shown in Fig.~\ref{Fig15}(a), Majorana modes can appear only in the interface between A and B. According to Eq.(\ref{theta}),  as long as $\theta \neq \theta_B$ and $\theta \neq \theta_A$, the condition 
\begin{eqnarray}
||\theta_A-\theta| - |\theta_B-\theta|| \lesssim 21^\circ-39^\circ \label{tricrystal}
\end{eqnarray}
is satisfied, there are Majorana zero-energy modes in the interface between A and B.
On the other hand, the supercurrent passing an junction interface between
superconductors $i$ and $j$ is given by $I=I_c \cos 2\theta_i \cos 2\theta_j$ (clean limit) or $I=I_c \cos 2(\theta_i + \theta_j)$ (dirty limit) \cite{Tsuei1, Rice, Tsuei2}, in which $\theta_i$ and $\theta_j$ are angles of the crystallographic axes(such as $(100)$) with respect to the junction interface. For the ring configuration shown in Fig.~\ref{Fig15}(a), in order that there is effectively a $\pi$-junction in the three junctions involved, we require $\cos^2 2\theta_C \cos 2\theta_A \cos 2\theta_B \cos 2(\theta-\theta_A) \cos 2(\theta-\theta_B)<0$ (clean limit) or $\cos 2(\theta_C+\theta_A) \cos 2(\theta_C+\theta_B) \cos 2(2\theta -\theta_A -\theta_B) <0$ (dirty limit). Hence to form a half-flux vortex in the ring,  in the clean limit, one requires $
\cos 2\theta_A \cos 2\theta_B \cos 2(\theta-\theta_A) \cos 2(\theta-\theta_B) <0$.
Therefore, it is sufficient to require one of angles, $\theta_A$,
$\theta_B$, $\theta-\theta_A$, and $\theta-\theta_B$ to be greater than $\pi/4$ and all the others to be less than $\pi/4$. There are two possibilities:  
\begin{eqnarray}
\text{either} \hspace{0.2cm} \pi/4<\theta_{A/B}<\pi/2 , \nonumber \\
\text{or} \hspace{0.2cm} \pi/4<\theta-\theta_{A/B}<\pi/2, \label{pi}
\end{eqnarray}
and all the other angles are less than $\pi/4$. As a result, the $\pi$-junction is either the junction between A (or B) and C or the junction between A and B.
For typical configurations of tricrystal experiments\cite{Tsuei1}, one has $\theta \simeq \theta_B$. The requirement of a ring to be a $\pi$-ring for the clean and dirty limits becomes the same and is given by
\begin{eqnarray}
\cos 2\theta_A \cos 2\theta_B \cos 2(\theta_A-\theta_B) <0. \label{tricrystal1}
\end{eqnarray}
On the other hand, Eq.(\ref{tricrystal}) implies $|\theta_A-\theta_B| \lesssim 21^\circ-39^\circ$, which, when combined with with Eq.(\ref{tricrystal1}), determines parameters that allow $\pi$ rings to host Majorana fermions. In Fig.~\ref{Fig15}(b), we summarize regions of parameters for $\pi$ rings and zero rings with Majorana fermions. It is seen that there are large parameter regions satisfying the requirement for hosting Majorana fermions. 
Therefore, one expects that the half-flux vortex trapped in a typical $\pi$-ring would jump in unit of two flux quanta in external magnetic fields.

\begin{figure}[tp]
\includegraphics[width=8.0cm]{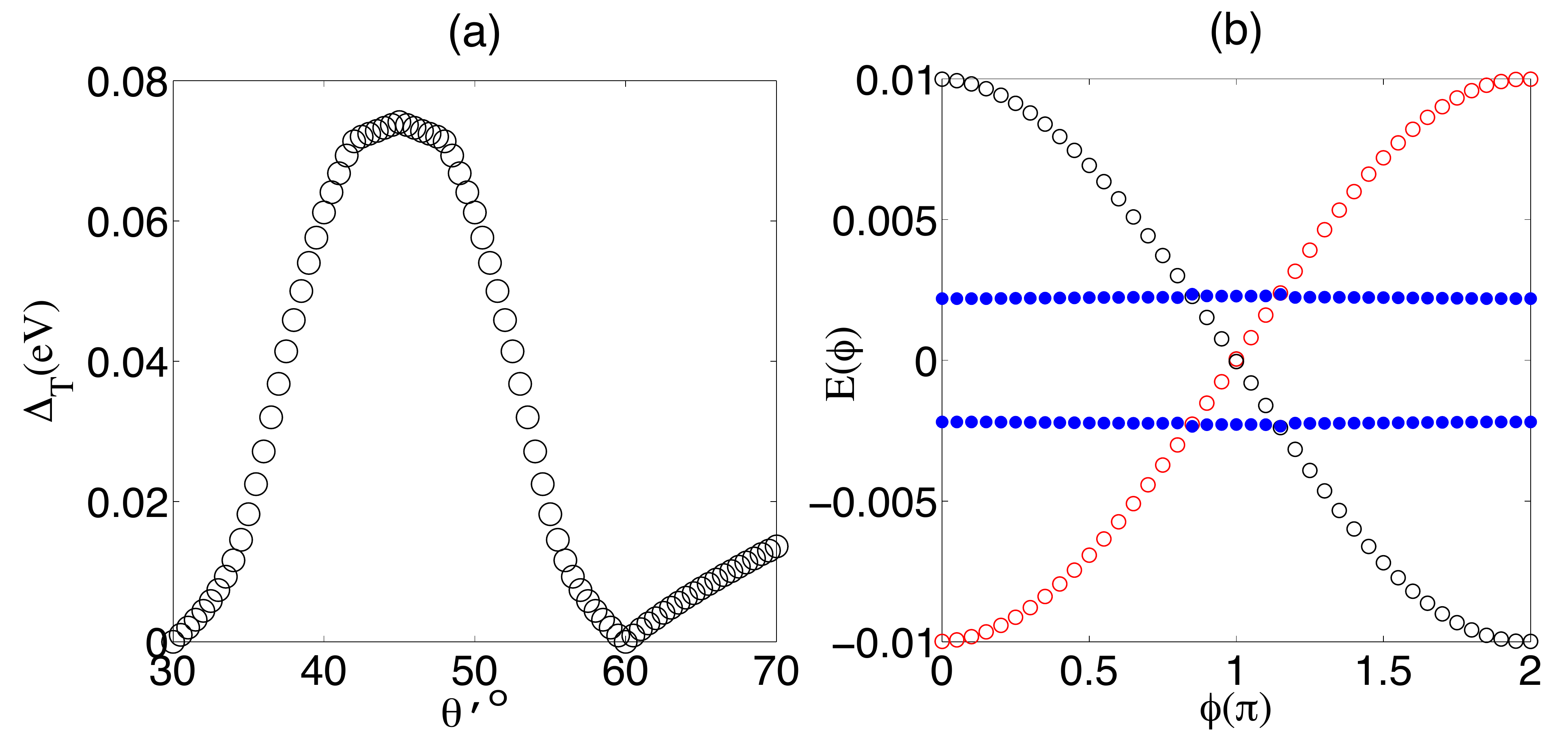} \caption{(a) Energy gap $\Delta_T$ due to the coupling of S and S$'$ in the SIS$'$ junction. Here $\theta$ is fixed to be $30^\circ$ and the energy for the quasi-particle of $S$ is chosen at the nodal point with the Fourier wavevector $p_y$. $\Delta_T$ is the minimum energy of quasi-particles in S$'$ with fixed $p_y$. $\Delta_T$ generally does not vanish except when S and S$'$ are the same or $\theta+\theta'=90^\circ$.  (b) Comparison of the enegy-phase relation due to quasi-particles (solid circle) and Majorana modes (open circle) localized at the SIS$'$ junction. The Josephson current due to quasi-particles is heavily suppressed.}\label{Fig16}
\end{figure} 

\subsection{Influence of quasi-particles and $k_z$ dispersion}
In this subsection, we examine stability of $4\pi$ periodicity and effects due to energy dispersion along $c$ axis.
Specifically, we examine whether the $4\pi$ periodicity is spoiled by the presence of quasi-particles.

We first note that the $4 \pi$ periodicity relies on the conservation of Fermion pairty. In the presence of quasi-particles, the fermion parity is no longer conserved\cite{Liang}.  However, if the energies of quasi-particles are gapped, the fermion parity is approximately conserved at low temperatures\cite{Liang}. As indicated in Fig.~\ref{Fig4}, the bulk superconducting states in realistic parameter regime are gapless. However, the SIS$'$ junction involves the coupling of the superconducting states in two different orientations. When two superconducting states couple, since the Fourier wavevector $p_y$ that is parallel to the junction is conserved, only quasi-particles with the same $p_y$ couple. Due to different orientations of S and S$'$ relative to the junction, $p_y$ of nodal points for S and S$'$ are not the same. As a result, quasi-particles at nodal points of S (or S$'$) couple to quasi-particles at finite energies of S$'$ (or S). Hence the resulting quasi-particles residing near the junction are gapful and there is an energy gap, $\Delta_T$, associated with each orientation of the tunneling junction. 
In Fig.\ref{Fig16}a, we examine $\Delta_T$ versus $\theta'$ by fixing $\theta=30^\circ$. Here the energy of the quasi-particle for S side is chosen at the nodal point with the Fourier wavevector $p_y$ and $\Delta_T$ is defined to be the minimum energy of quasi-particles in S$'$ with fixed $p_y$. It is seen that except for special orientations when S and S$'$ are the same or $\theta+\theta'=90^\circ$ , $\Delta_T$ generally does not vanish. The parity conservation thus holds for quite general orientations of S and S$'$. In addition to the protection of the pairty conservation, the momentum mismatch also suppresses the quasi-particle contribution. This is illustrated in Fig.~\ref{Fig16}b. It is seen that $E(\phi)$ for the contribution of quasi-particles is much more flat in comparison to that due to Majorana modes. Hence the Josephson current ($\frac{d E(\phi)}{d \phi}$) due to quasi-particles is heavily suppressed and leads to little effects. 

Finally, we discuss effects of the energy dispersion ($k_z$) along $c$-axis. It is generally accepted that high-$T_c$ cuprates are approximately two dimensional with weak dispersion along $c$-axis. Therefore, results derived in the above are approximately correct. For high-$T_c$ cuprates with one CuO$_2$ per unit cell, the energy dispersion along $c$-axis can be approximated by\cite{kz} $\Delta \xi_k = -2 t_z \cos (k_z/2) (\cos k_x - \cos k_y)^2 \cos k_x/2 \cos k_y/2$ with $t_z \sim 0.1 t$. Following the discussion on Eq.(\ref{spectra}), for a given $k_z$, nodal points are determined by $\xi_k+\Delta \xi_k=0$. In other words, $\Delta \xi_k$ changes the Fermi surface. For realistic parameters, $|\alpha_R| > \Delta_p$, nodal points are along axes $k_y = \pm k_x$ on which $\Delta \xi_k=0$ is satisfied. Hence nodal points have no $k_z$ dispersion. Therefore, all results based on projection of nodal points (including $4 \pi$ periodicity) are the same except that there are more Majorana edge modes due to the $k_z$ dispersion. On the other hand, for $\Delta_p > |\alpha_R|$,  the $k_z$ dispersion changes the Fermi energy by 10\% for a give $k_z$. As a result, there is no qualitative change on the dispersive Majorana modes except for the corresponding 1\% change of the range for the existence of the dispersive Majorana modes. 

\section{Discussion and Conclusion}
\label{conclusion}

In conclusion, we have explored effects of the Rashba spin-orbit interaction on the tunneling spectroscopy of 
high-$T_c$ cuprate superconductors. 
The mean-field phase diagram in the large-U limit of Hubbard model is established.
It is shown that due to the Dzyaloshinskii-Moriya and spin dipole-dipole interactions induced by the spin-orbit interaction, the majority regime in the phase diagram is gapless with pairing symmetry of superconductivity being $p+d$-wave. Furthermore, the gap function $d$-vector for $p$-wave superconductivity is not aligned with the internal magnetic field of the spin-orbit interaction. 
As the spin-orbit interaction is turned on, we find that the ground state undergoes a phase transition to a topological gapless phase with each nodal point originated from pure $d$-wave being split into two stable nodal points characterized by the symmetry class DIII. Due to the splitting nodal structure, zero-energy Majorana modes always exist for any interfaces that are not exactly in $(100)$ or $(010)$ directions.
In addition, due to non-aligned $d$-vector, as the $p$-wave amplitude further increases beyond the Rashba spin-orbit interaction, a transition of nodal point configuration occurs.  The ground state is still gapless. However, the Majorana flat band near $p_y=0$ in $(110)$ edge becomes the dispersive Majorana edge modes. 

In addition to the mean-field phase diagram, the tunneling conductances for the NIS and SIS$'$ junctions are also computed.
Our results indicate that due to the presence of dispersive Majorana modes, a small plateau with shoulders near the zero bias peak would be induced in the tunneling spectrum. This small plateau with shoulders is a result due to the Dirac cone of dispersive Majorana modes at the edge. The appearance of the Dirac cone at $p_y=0$ shown in Fig.~\ref{Fig11} is a manifestation of $p$-wave order parameter and results from competition between the $d$-wave and $p$-wave pairing symmetries. When the $d$-wave dominates for $|\alpha_R |> |\Delta_p|$, there are only flat bands at the edge, exhibiting a single ZBCP in the tunneling spectrum. Only when $|\alpha_R | < |\Delta_p|$, the Dirac cone near $p_y=0$ emerges and results in the additional feature that exhibits the feature of a small plateau with shoulders. 

The existence of the Dirac cone at the edge requires particular symmetries. General perturbations such as disorders do not respect these symmetries and would not yield the feature of a small plateau with shoulders in the tunneling spectrum\cite{Lee}. Hence the corresponding feature can not be generated by disorders. The feature of a zero-bias conductance peak sitting on a small plateau with shoulders has been frequently observed in tunneling spectrum of $(110)$ edge during the past\cite{Tsuei, shoulder}.  Even though these evidences may not be the unique signature to conclude the existence of dispersive Majorana fermions and further evidence from other measurements is in order, they show supportive evidence for dispersive Majorana modes and indicate the importance of the spin-orbit interaction in high-$T_c$ cuprates.

In addition, we find that the overlap of dispersionless Majorana flat-bands in a Josephson junction always results in 4$\pi$-periodicity in Josephson effect. In particular, 
for typical configurations of $\pi$-ring in tricrystal experiments that can trap a half-flux vortex, we find that as long as the difference of the orientations for $a$ axes of two crystals involved in a junction is within $21^\circ-39^\circ$, the junction always exhibit 4$\pi$-periodicity in Josephson effect. 
For typical configurations of tricrystal experiments, one has $\theta_A=30^\circ$, $\theta=56.5^\circ$, and $\theta_B=56.5^\circ$\cite{Tsuei1}. The edge of crystal B in facing A is along $(010)$ direction. If the orientation of interfaces is perfect, since there is no edge state in $(010)$ edge, one expects that there is no Majorana fermion trapped in the $\pi$-ring. However, in real experiments, there may exist deviations in orientations of edges. Any deviation of the edge for crystal B from the $(010)$ direction would satisfy Eq.(\ref{tricrystal}). This would result in trapped Majorana fermions between A and B so that
the trapped flux in the $\pi$-ring jumps in unit of two flux quanta in externally applied magnetic fields. This would be consistent with the experimental observation that $3/2$-flux quanta has not been observed in tricrystal experiments\cite{Tsuei1}. Therefore, the above analyses indicate that Majorana fermions may have been already observed in tunneling experiments on high-$T_c$ cuprates. 

While so far in this work we only consider results based on the mean-field theory, we expect our results
are robust in the presence of  correlation effects as long as the symmetry of the system is not changed. Even though our results show agreement with past experimental observations, definite confirmation of Majorana fermions in high-$T_c$ cuprates requires further experimental studies. Nonetheless, our results offer important signatures to look for in future experiments. In particular, the setup configuration for tricrystal experiments would offer a unique way to hold Majorana fermions in high-$T_c$ cuprates. All the signatures we find are crucial for successfully searching Majorana fermions and are left for future experimental confirmations.

\begin{acknowledgments}
We thank Profs. John Chi and Chung-Hou Chung for useful discussions. This work was
supported by the Ministry of Science and Technology (MoST) of Taiwan.
\end{acknowledgments}


\end{document}